\documentclass[10pt,a4paper]{article}
\usepackage{amsmath, amsthm,amssymb,bm, tikz, breqn, soul}
\usepackage[sort&compress]{natbib}
\usepackage{mathptmx}
\usepackage{graphicx} 
\usepackage{lastpage}
\usepackage{float}
\usepackage{fancyhdr}
\usepackage{authblk}
\usepackage{upgreek}
\usepackage{charter}
\usepackage[colorlinks=true,linkcolor=blue,citecolor=blue
]{hyperref}
\usetikzlibrary{math}
\newcommand{\mg}[1]{\mathring{#1}}
\newcommand{\mb}[1]{\mathbf{#1}}
\theoremstyle{remark}
\newtheorem*{remark}{Remark}
\newcommand{\fo}{\mg{\mb{f}}}
\newcommand{\Xo}{\mg{X}}
\newcommand{\Deltao}{\mg{\Delta}}
\newcommand{\Fo}{\mg{\mb{F}}}

\begin{document}

\title{Instabilities in Colloidal Crystals on Fluid Membranes}

\author[1]{Sanjay Dharmavaram}
\affil[1]{Department of Mathematics and Statistics, Bucknell University, Lewisburg, PA 17837;
Email: {sd045@bucknell.edu}}

\author[2]{Basant Lal Sharma}
\affil[2]{Basant Lal Sharma, Department of Mechanical Engineering, Indian Institute of Technology Kanpur, Kanpur, 208016 UP, India;
Email: {bls@iitk.ac.in}}

\date{\today}
\maketitle 

\begin{abstract}{
The complex physics of self-assembly in colloidal crystals on deformable interfaces and surfaces poses interesting possibilities for the designability and synthesis of next-generation metamaterials. The goal of this article is to characterize instabilities arising in colloidal crystals assembled on fluid membranes.  The colloidal particles are modeled as pair-wise interacting point particles, constrained to lie on a fluid membrane and yet free to reorganize, and the membrane's elastic energy is modeled via the Helfrich energy. We find that when a collection of particles is arranged on a planar membrane in some regular fashion -- such as periodic lattice -- then the regular configuration admits bifurcations to non-planar configurations. Using the Bloch-wave anstaz for the mode of instabilities, we present a parameteric analysis of the boundary between the stable and unstable regimes. We find that instabilities can occur through two distinct kinds of modes, when the parameters belong in certain physically interesting regimes, referred to as long-wavenumber modes ($L$ modes) and short-wavenumber modes ($S$ modes) in the article. We discuss some connections between these results and recent experiments, as well as the open problem of budding in biomembranes.}
\end{abstract}

\section{Introduction}
\label{sec:introduction}

Two-dimensional arrays of colloid nanoparticles deposited on interfaces are playing a crucial role in the design of next-generation metamaterials. Applications range from energy-harvesting  \citep{sargent2012colloidal, kramer2014colloidal, emin2011colloidal}, biomedical  \citep{dinsmore2002colloidosomes}, to photonic materials  \citep{cai2021colloidal} and sensors, and structural color \citep{liao2019multiresponsive, gu2003structural, liu2019self}. One of the striking features of colloidal particles is their ability to spontaneously assemble into clusters, often exhibiting a crystalline arrangement. This self-assembly process can, to a certain extent, be directed, for instance, by altering the nature of interparticle interactions, changing the pH of their environment, etc. This high degree of tunability can be leveraged for designing materials with tailored properties.

Colloidal crystals are also of interest as model systems for answering fundamental questions about material properties. Colloidal crystals grown on curved substrates have been used to understand curvature-induced defects and the role of geometrical frustration  \citep{sachdev1984crystalline, dinsmore1998hard, bowick2000interacting, vitelli2006crystallography, irvine2010pleats, irvine2012fractionalization, garcia2013crystallization, meng2014elastic, manoharan2015colloidal,  brojan2015wrinkling,  stoop2018defect}. They are also used as mesoscopic models for understanding the mechanics of assembly of viral capsids~ \citep{bruinsma2003viral,luque2012physics, wagner2015robust, perotti2016useful}.

When colloidal crystals assemble on deformable substrates, novel forms of instabilities are observed~ \citep{huang2007capillary, stafford2004buckling}. When Polystyrene (PS) nanoparticles, deposited on a polydimethylsiloxane (PDMS) substrate, are subjected to compressive stresses,  buckling instability of the substrate into a sinusoidal pattern is observed \citep{kusters2019actin, gurmessa2013onset, gurmessa2017localization}. 

\begin{figure}[h!]
    \centering
    \caption{Fluorescent confocal microscopy images of PNIPAM microgel particles assembled on DMPC lipid membrane. The figure is adapted from Fig. 3 of the work of Wang, et. al.   \citep{wang2019assembling}}
    \label{fig:Wangs experiment}
\end{figure}
Recently, Wang, et. al  \citep{wang2019assembling} have successfully synthesized microgel particles adsorbed on lipid membranes (lipogels). Fig.~\ref{fig:Wangs experiment}, adapted from their work, shows fluorescent confocal microscopy images of N-isopropylacrylamide (PNIPAM) microgel particles assembled in a regular arrangement on 1,2-dimyristoyl-sn-glycero-3-phosphocholine (DMPC) lipid membrane. Wang, et. al  \citep{wang2019assembling} observe that by controlling the fluidity of the lipid membrane, the self-assembly can be controlled. 

Most theoretical and computational studies have so far focused on the assembly of colloidal particles on rigid substrates \citep{meng2014elastic, perotti2016useful, webb2025curvedspacesim}. 
In fact, the results of mathematical analysis for two dimensional lattices, where particles interact with a (generalized) Lennard-Jones potential, have been exhaustively studied by B\'{e}termin, et. al \citep{betermin2015minimization}.
The role of long-range and cooperative interactions between particles and substrate elasticity has been explored less in the existing literature.  
The questions regarding the influence of substrate elasticity on particles' arrangement and particle/substrate interaction motivate further research.

The primary goal of our study, partly motivated by the researches of Wang et. al \citep{wang2019assembling} and Bekele et. al \citep{gurmessa2013onset, gurmessa2017localization}, is to characterize instabilities that arise in colloidal crystals on a planar fluid membrane. Specifically, our focus is on instabilities that induce the system to assume non-planar configurations. 

This article is organized as follows. In Sec.~\ref{sec:formulation}, we present a discrete-continuum model for the system. The membrane is represented as a two-dimensional (2D) surface with an elastic energy given by the Helfrich energy. The particles are modeled as point particles interacting via a pair potential. We present the equilibrium equations derived from variational techniques. In Sec.~\ref{sec: local instability analysis}, using the Bloch ansatz, we present the linear instability analysis and derive the condition under which a flat state (planar configuration) loses stability to out-of-plane modes. We present the results of our analysis in Sec.~\ref{sec:results} and explore connection to experiments in Sec.~\ref{sec:discussion}.

\section{Formulation}
\label{sec:formulation}
\begin{figure}[h!]
    \centering
    \includegraphics[width=\linewidth]{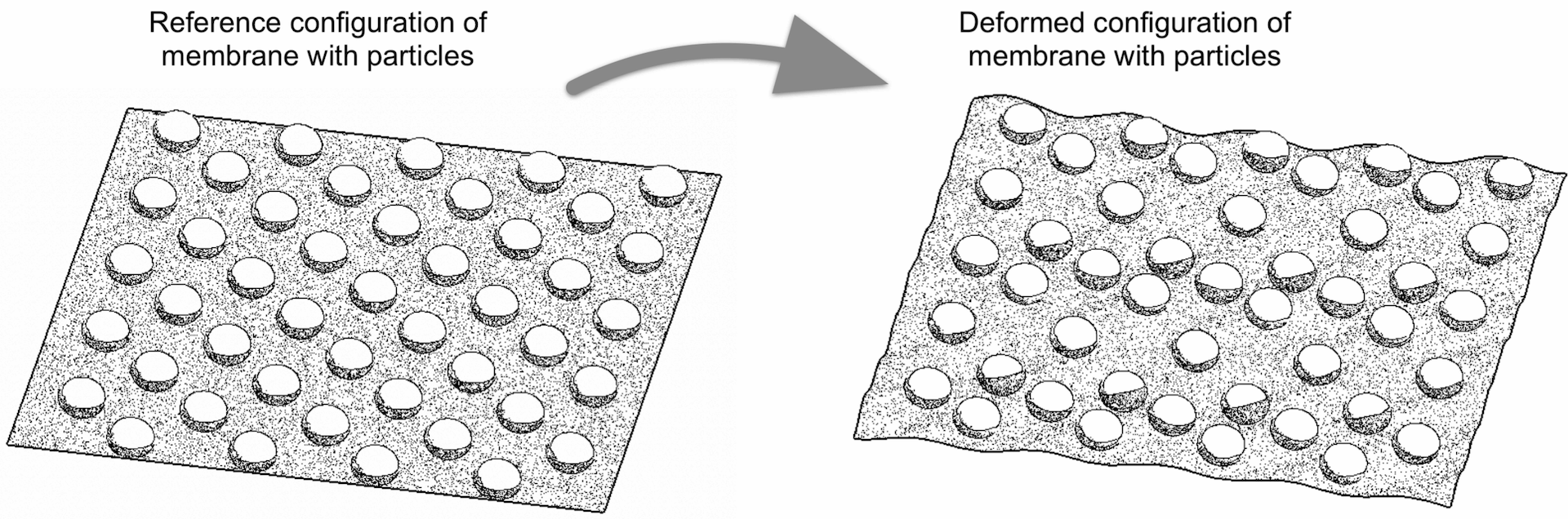}
    \caption{A schematic showing reference and deformed configurations of the membrane-particle system.}
    \label{fig:schematic}
\end{figure}

We consider a membrane-particle system consisting of an infinite fluid membrane, geometrically modeled as a surface, embedded with colloidal particles, as shown in the schematic Fig.~\ref{fig:schematic}. 

We physically model the fluid membranes using the Helfrich model  \citep{helfrich1973elastic}, and the colloidal particles as point particles, interacting with each other via a pair potential.  Since a lipid membrane (under most physiological conditions) behaves like a 2D fluid and permits negligible amount of resistance for in-plane shearing, we assume that the particles, even though they are adhered to the membrane, are free to float around in the plane of the membrane. It has been observed in experiments \citep{wang2019assembling} that the fluidity of the membrane aids in the ordering of the particles into a regular crystalline arrangement even though the particles are not allowed to leave the surface of the membrane. In particular, an out-of-plane motion of the particles induces an out-of-plane deformation in the membrane which eventually involves a balance between the un-equilbriated force due to out-of-plane motion of the particles and the force due to bending elasticity of the membrane.

In our formulation, to anchor the particles on the membrane without having to resort to any explicit constraints, we employ the \emph{Lagrangian formulation} as developed in  \citep{dharmavaram2020lagrangian, dharmavaram2022lagrangian}. In this formulation, we parameterize the position of particles using their reference position $\mathbf{X}_i\in\mathbb{R}^2$, where the index $i$ runs over the set of total $N$ particles (in this article, $\mathbb{R}^2$ denotes 2D plane and $\mathbb{R}^3$ denotes three-dimensional space). If the surface were to undergo any deformation in the three-dimensional space $\mathbb{R}^3$, then the current (actual) position of the particles on the membrane is given by
\begin{equation}
\mb{f}_i:=\mb{f}(\mb{X}_i), \quad i=1, \dotsc, N,
\label{deffXi}
\end{equation}
where  $\mb{f}:\mathbb{R}^2\to\omega\subset \mathbb{R}^3$ is the deformation map for the lipid membrane surface $\omega$. 

As mentioned above, the elastic energy, $\mathcal{E}$, of the system is a combination of the Helfrich energy and the total pair-wise interaction energy of the particles.  That is,
\begin{equation}
    \mathcal{E}[\mb{f}, \{\mb{X}_i\}] := \int_{\omega}\kappa H^2 \;da +
    \frac{1}{2}\sum_{\substack{i, j=1\\i\neq j}}^{N}V(\lVert\mb{f}_i-\mb{f}_j\rVert),
    \label{eq:energy}
\end{equation}
where 
 $\kappa$ denotes the bending stiffness, $H$ denotes the mean curvature of the membrane, and 
 $V(r)$ is the assumed  pair potential for a given radial distance  $r$ between any two particles. 
 The dependence of the energy functional on the degrees-of-freedom $\mathbf{f}$ and $\mathbf{X}_i$ ($1\leq i\leq N$) is explicitly indicated in \eqref{eq:energy}.
 
 In \eqref{eq:energy}, since we are assuming an infinite membrane (i.e., no boundary), the Gaussian curvature term, typically included in the Helfrich energy has been dropped due to the Gauss-Bonnet theorem  \citep{carmo1992riemannian}. 
In this work, we restrict our considerations to a periodic system of particles, and the sum in \eqref{eq:energy} being over all the particles amounts to $N=\infty$. Note that, implicit in the form of the elastic energy \eqref{eq:energy} is the assumption of a weak adhesion limit, where the colloidal particles remain embedded (from one side) on the physical membrane without significantly deforming the membrane through adhesive interactions. 
For sufficiently high surface tension, this assumption is a reasonable approximation, as noted in the experiments of Wang et. al \citep{wang2019assembling}. 
Furthermore, we assume that the pair-wise interaction energy $V$, of the particles, has the form of a generalized Lennard-Jones (LJ) potential \citep{betermin2015minimization}, expressed as,
 \begin{equation}
    V(r) = \frac{\epsilon}{\sigma^n-\sigma^m} \left[\left(\frac{\sigma}{r}\right)^m -\left(\frac{\sigma}{r}\right)^n \right],
    \label{eq:LJ}
\end{equation}
where $m>n>0$, and   $\epsilon$ is the minimum value of $V$ attained at an equilibrium distance $r=r_e:= \sigma\sqrt[m-n]{m/n}$. For the standard LJ potential, $m=12$ and $n=6$. For \emph{macroscopic} bodies, in particular spheres of finite radius, the Van der Waals interaction is also given by a 7-1 potential  \citep{butt2023physics, hamaker1937london, henderson1997expression}, that is \eqref{eq:LJ} with the choice $m=7$ and $n=1$.

Since fluid membranes such as lipid membranes have a area stretching modulus that is significantly larger than the bending modulus, they are commonly approximated to be area-preserving \citep{jenkinsRBC}. Accordingly, we impose the local area constraint ${da}/{dA} \equiv 1$, which is equivalent to
\begin{equation}
J:=\frac{\sqrt{g}}{\sqrt{G}}\equiv 1,
 \label{eq:j=1 constr final}
\end{equation}
where $g$ and $G$ are the determinants of the metric tensor of the current (deformed) surface $\omega$ and reference surface $\mathbb{R}^2$.

Using the standard techniques of the calculus of variations,  we take the first variation of the energy functional $\mathcal{E}[\mb{f}, \{\mb{X}_i\}]$ \eqref{eq:energy} with respect to the degrees-of-freedom,  $\mathbf{f}$ and $\mathbf{X}_i$ (see Supplementary Material 
 for details), as related by \eqref{deffXi} for the particle-membrane system, to obtain the following equilibrium (Euler-Lagrange) equations:

\begin{subequations}
\begin{equation}
    \kappa\Delta H +2 H \kappa(H^2-K)-2H\gamma + \sum_{i=1}^N \mathbf{n}\cdot\mb{F}_{T,i}\delta(\mb{X}-\mb{X}_i)=0,
    \label{eq:eqlbrmnrml simp}
\end{equation}

\begin{equation}
    \mb{F}_{T,i}\cdot\mb{f}_{,\alpha}(\mb{X}_i)=0,\, \forall i=1,\cdots,N\text{ and }\alpha=1,2,
    \label{eq:eqlbrm part simp}
\end{equation}
where:\\
$\Delta(\cdot)$ is the Laplace-Beltrami operator on the surface $\omega$, \\
$\mathbf{n}$ is the unit normal vector field on the surface $\omega$, \\
$\gamma$ is the surface tension in the membrane is imposed as a Lagrange multiplier to enforce the area constraint \eqref{eq:j=1 constr final},  \\
$\delta$ is the Dirac delta distribution on $\mathbb{R}^2$, \\
and
 $\mathbf{F}_{T,i}$ is the net force acting on the particle $i$ due to all other particles, viz.,
\begin{equation}
    \mb{F}_{T,i}:=\sum_{j=1,j\neq i}^N \mb{F}_{ij}, 
\label{eq:F_Ti}
\end{equation}
where
\begin{equation}
    \mb{F}_{ij}:=V'(r_{ij})\frac{\mb{f}_i-\mb{f}_j}{r_{ij}},\text{ with } r_{ij}:=\lVert\mb{f}_i-\mb{f}_j\rVert.
\label{eq:Fij}
\end{equation}
\label{eqs:equilibrium}
\end{subequations}
Equation \eqref{eq:eqlbrmnrml simp} is the  moment balance for the membrane, and \eqref{eq:eqlbrm part simp} is the force balance of the particles. It can be shown that the latter implies that the Lagrange multiplier $\gamma$ enforcing the area constraint is a constant (see Supplementary Material
for details); thus, the surface tension in the membrane is uniform.

\begin{remark}
In what follows, we non-dimensionalize the equilibrium equations  \eqref{eq:eqlbrmnrml simp}-\eqref{eq:Fij} by choosing the energy scale in the units of $\epsilon$ (the minimum value of $V$) and the length scale in units of $r_e$ (the minimizer of $V$). This introduces the two non-dimensional quantities: $\hat{\kappa}=\kappa/\epsilon$ and $\hat{\gamma} = r_e^2\gamma/\epsilon$. For notational convenience, in the following two sections, we ignore the decoration $\hat{\cdot}$ and write $\hat{\kappa}$ as $\kappa$ and $\hat{\gamma}$ as $\gamma$, with the understanding that $\epsilon=1$ and $r_e=1$. In Sec.~\ref{sec:discussion}, where we compare our analysis to experimental data, we revert to $\hat\kappa$ and $\hat\gamma$.
\end{remark}

\section{Local instability analysis of flat state using Bloch modes}
\label{sec: local instability analysis}

In this section, we undertake a local instability analysis of \eqref{eq:eqlbrmnrml simp}--\eqref{eq:Fij} to determine conditions under which a flat state with particles arranged in a regular lattice loses stability to non-planar states. In this context, one of the main questions we explore in this article is the role of coupling between the particles and the deformability of the substrate. Specifically, we aim to investigate how inter-particle interactions induce an out-of-plane deformation on the membrane. We, therefore, do not consider in-plane symmetry breaking where the particles rearrange in the plane of the membrane without causing any out-of-plane displacement. A careful stability analysis of this type for various possible 2D lattices of generalized Lennard-Jones particles has already been presented by B\'{e}termin, et. al  \citep{betermin2015minimization}.

\subsection{Flat state as trivial solution}
\label{ssec:trivialsol}

\begin{figure}[h!]
    \centering
\begin{tikzpicture}
\begin{scope}
\tikzmath{\sc = 0.20;}
\foreach \m in {-1,0,...,2}{ 
    \foreach \n in {-1,0,...,2}{
    \node[draw,circle,inner sep=1pt,fill] at ({\m*1+cos(60)*\n},{sin(60)*\n}) {}; 
    }
}
\draw [blue,->] (0,0) -- (1,0) node[right] {$\mg{r}\mathbf{a}_1$};
\draw [blue,->] (0,0) -- ({cos(60)},{sin(60)}) node[right] {$\mg{r}\mathbf{a}_2$};

\draw [red,->] (0,0) -- ({\sc*4*pi/sqrt(3)*cos(60-90)}, {\sc*4*pi/sqrt(3)*sin(60-90)}) node[below] {$\mathbf{G}_1$};
\draw [red,->] (0,0) -- ({\sc*4*pi/sqrt(3)*0}, {\sc*4*pi/sqrt(3)*1}) node[left] {$\mathbf{G}_2$};
\fill[gray, opacity=0.6] (0,0) -- (1,0) -- ({1+cos(60}, {sin(60}) -- ({cos(60)},{sin(60)}) -- cycle;
\end{scope}
\end{tikzpicture}
    \caption{Schematic showing triangular lattice arrangement of particles. The basis vectors are shown in blue and the reciprocal basis in red. The gray region highlights the unit cell.}
    \label{fig:triangular lattice}
\end{figure}
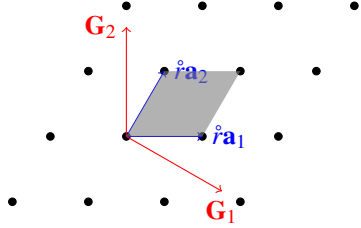

It is clear that, due to symmetry, a flat state (planar configuration) of particles arranged in a 2D Bravais lattice is an equilibrium configuration of the system, referred as the trivial solution of the equations of equilibrium of membrane-particle system. The stability of the states of such possible 2D lattices is nontrivial. As shown in the works of B\'{e}termin and others  \citep{betermin2015minimization,travvenec2019two}, a  triangular or square Bravais lattice is a stable equilibrium state for a range of nearest neighbor spacings $\mg{r}$. In such a lattice, the equilibrium position of the particle $i$ in the reference configuration is given by 
\begin{equation}
    \mg{\mb{X}}_i = \mg{r}[m_i\mb{a}_1 + n_i\mb{a}_2],
    \label{eq:lattice pos}
\end{equation}
where  $m_i,n_i\in \mathbb{Z}$ and for the case of a triangular lattice, the basis vectors are $\mb{a}_1=\langle 1,0\rangle, \mb{a}_2=\langle \frac{1}{2},\frac{\sqrt{3}}{2}\rangle$, while in the case of a square lattice, $\mb{a}_1=\langle 1,0\rangle, \mb{a}_2=\langle 0,1\rangle$. 
Here we have denoted the Cartesian coordinates of a point $\mathbf{X}$ in $\mathbb{R}^2$ by the symbols $(X_1,X_2)$, while $\mb{e}_1$, $\mb{e}_2$ denote the  basis vectors in $X_1, X_2$ directions, respectively.
We define
\begin{equation}
\mg{r}:=\left\{ 
\begin{array}{cc}
\ell\sqrt{2/\sqrt{3}} & \text{ triangular lattice}\\
\ell & \text{ square lattice}
\end{array}
\right.
\label{eq:r0}
\end{equation}
where, $\ell^2$ is the area per unit cell (area of gray region in the schematic Fig.~\ref{fig:schematic}). In this article, we focus our attention largely on the triangular lattice (we discuss very briefly the case of square lattice later). A schematic for a triangular lattice is shown in Fig.~\eqref{fig:triangular lattice}, where the dots represent  particle location.

The parameter $\ell$ in \eqref{eq:r0} can be identified with $A$ in the works of B\'{e}termin, et. al \citep{betermin2015minimization}.  It should be noted that even though their results were derived for planar lattices without any fluid membrane,  in the presence of a membrane, these configurations are still equilibrium states for any value of  $\kappa$ and $\gamma$. This claim can be verified by noting that for a planar membrane, the deformation mapping is 
\begin{equation}
\mg{\mb{f}}(\mathbf{X}) =\mg{\mb{f}}(X_1,X_2) = X_1\mathbf{e}_1 + X_2\mathbf{e}_2,
\label{eq:trivial f}
\end{equation}
whence 
\begin{equation}
H(X_1,X_2)=K(X_1,X_2)\equiv 0, J(X_1,X_2)\equiv 1, \mb{n}=\mb{e}_3 \text{(z-axis)}.
\end{equation}
Also, since the particles are in the $X_1$-$X_2$ plane, $\mb{F}_{ij}\cdot{\mb{n}}=0$. Therefore, for any $\kappa$ \eqref{eq:eqlbrm part simp}, \eqref{eq:j=1 constr final} are trivially satisfied. Furthermore. $\mg{\mathbf{F}}_{T,i}=\mathbf{0}$ (net force acting on particle $i$ due to all other particles in the $X_1$ and $X_2$ directions) which follows from symmetry considerations.

As discussed above, \eqref{eq:lattice pos}, \eqref{eq:r0}, and \eqref{eq:trivial f} form a \emph{trivial solution} of the problem. In above, and in the rest of the article, we distinguish the quantities associated with the trivial flat state by placing a circle on top of the appropriate quantity.
In particular, the value of ${\gamma}$ for a trivial state is denoted as $\mg{\gamma}.$ Physically, this represents surface tension. In a finite membrane, this value will be determined by the tensile force applied at the boundary. In the case of slightly curved membranes, i.e., when $\mg{r}$ is significantly smaller than the radius of curvature of the membrane, $\gamma$ is set by the osmotic pressure on the membrane.
Note that, in the flat case, the problem is degenerate as any arbitrary value for $\mg\gamma$ is allowed. We therefore treat it as a parameter of the problem. 
We explore the existence of bifurcating solutions from such a trivial state as described thus far. We next derive the linearization of \eqref{eqs:equilibrium} about the trivial state.

\subsection{Linearization}
\label{ssec:linearization}

Suppose that the perturbation of the surface, away from the flat configuration \eqref{eq:trivial f}, is described by
\begin{equation}
    \mathbf{f}(\mathbf{X})= \fo(\mathbf{X})+{z(\mathbf{X})\mathbf{e}_3},  \quad z:\mathbb{R}^2\to\mathbb{R},
    \label{fpertbn}
\end{equation}
and that the perturbed particles' positions, away from \eqref{eq:lattice pos}, are give by 
\begin{equation}
    \mb{X}_i=\mg{\mb{X}}_i + {\bm\alpha}_i
    \label{Xipertbn}
\end{equation}
where $z(\mathbf{X})$ the Monge parametrization of the surface, and $\bm{\alpha}_i$ is the variation in particle $i$'s reference position. Let us denote the perturbation in $\mg{\gamma}$ as $\tau$. That is,
\begin{equation}
    \gamma=\mg{\gamma}+\tau.
    \label{gpertbn}
\end{equation}
We assume that
$(z, \bm{\alpha}_i, \tau) = O(\xi)$, 
where $\xi$ is a small real number (representing the amplitude of perturbations).

For particle $i$ and $j$, let us define:
\begin{subequations}
\begin{equation}
    \mg{\mathbf{f}}_i:=\mg{\mathbf{f}}(\mg{\mathbf{X}}_i),\;\ell_{ij} := ||\fo_i - \fo_j||, 
\label{eq:f_i, ellij}
\end{equation}
\begin{equation}
    \bm{\alpha}_{ij}:=\bm{\alpha}_i-\bm{\alpha}_j,
\label{eq:alpha_ij}
\end{equation}
\begin{equation}
    \mathbf{\Xo}_{ij}:=\mathbf{\Xo}_i-\mathbf{\Xo}_j,\;z_{ij} := z(\mb{\Xo}_i)-z(\mb{\Xo}_j).
    \label{eq:Xij}
\end{equation}
\label{eqs:misc defs}
\end{subequations}

It follows  (see Supplementary Material
 for details) that the linearization of \eqref{eq:eqlbrmnrml simp} is given by,
\begin{equation}
   \kappa\Deltao^2 z-2\mg{\gamma}\Deltao z  + 2 \sum_{\substack{i, j\\j\neq i}} \mg{A}_{ij} \Big(z_{ij}-\nabla z(\mathbf{\Xo}_{i}) \cdot \mathbf{\Xo}_{ij}\Big)\delta(\mathbf{X}-\mg{\mathbf{X}}_i) = 0,
   \label{eq:flineqn}
\end{equation}
where $\mg{A}_{ij}=V'(\ell_{ij})/\ell_{ij}$, $\mg{\Delta}(\cdot)=(\cdot)_{X_1X_1} + (\cdot)_{X_2X_2}$ is the Laplacian-Beltrami operator on the flat state, and $\mg{\Delta}^2(\cdot)$ is the corresponding biharmonic operator. Using the fact that $\Fo_{T,i}=\mb{0}$ on the trivial state, the  linearization of \eqref{eq:eqlbrm part simp} reads:
\begin{equation}
    \sum_{\substack{j\neq i}}\left[\frac{\left(V''(\ell_{ij}) -\mg{A}_{ij}\right)}{\ell_{ij}^2}\mathbf{\Xo}_{ij}\otimes\mathbf{\Xo}_{ij} + \mg{A}_{ij}\mathbf{I}\right]\bm{\alpha}_{ij} = \mb{0},\text{ for each } i .
   \label{eq:Xilineqn}
\end{equation}
Here $\otimes$ represents the tensor product and $\mathbf{I}$ is the $2\times 2$ identity operator.

Since the trivial state is flat, the linearization of the incompressibility constraint \eqref{eq:j=1 constr final} is automatically satisfied (see Supplementary Material
 for details). 

We see from \eqref{eq:Xilineqn} that the equation governing the perturbations of the position of the particles in the reference configuration ($\bm{\alpha}_i$) is uncoupled from that of the membrane, i.e., \eqref{eq:flineqn}. Thus, at linear order, the 3D configuration of the surface is not associated with a rearrangement of the in-plane configuration of the particles. This, however, does not mean that the particles have no influence on the 3D shape. This can be seen from the third term of \eqref{eq:flineqn} that depends on the particle interactions. In fact, an alternative approach, by minimizing particle interaction energy to determine stability of planar lattice configurations, has already been explored in the works of B\'{e}termin et al  \citep{betermin2015minimization}. Our focus in this article, on the other hand, is to understand \emph{how the particles can induce shape changes in the membrane}. Therefore, we will not explore the nontrivial solutions of \eqref{eq:Xilineqn} and instead focus on the nontrivial solutions of \eqref{eq:flineqn}. 

\begin{remark}
Note that the linearization of $\gamma$, viz., $\tau$ remains arbitrary as the linearization of $J=1$ is trivially satisfied. Since $\tau$ is small and  is of $O(\xi)$, it is necessarily zero. 
\end{remark}

To determine bifurcating solutions from the trivial state, we seek a certain type of nontrivial solutions of \eqref{eq:flineqn}. We do this using a Bloch ansatz for $z$ as follows. 

\subsection{Bloch ansatz}

For the lattice considered in Fig.~\eqref{fig:triangular lattice}, we denote the reciprocal basis with $\mathbf{G}_1$ and $\mathbf{G}_2$ (also shown in the same Fig.~ \ref{fig:triangular lattice}). Let 
\begin{equation}
\mathcal{W}:=\{\mu_1 \mathbf{a}_1+\mu_2 \mathbf{a}_2: \mu_1,\mu_2\in[-1/2,1/2]\}
\label{eq:defWg}
\end{equation}
denote the Wigner-Seitz cell and 
\begin{equation}
\mathcal{B}:=\{\beta_1 \mathbf{G}_1+\beta_2 \mathbf{G}_2: \beta_1,\beta_2\in[-1/2,1/2]\}
\label{eq:defBz}
\end{equation}
denote the first Brillouin zone. Then, for any $\mathbf{X}\in \mathcal{W}$, the Bloch ansatz for a solution of \eqref{eq:flineqn} is
\begin{equation}
z(\mathbf{X})=\int_{\mathcal{B}}z(\mathbf{K},\mathbf{X})d\mathbf{K},
\label{eq:integral z}
\end{equation}
where, for all $\mathbf{K}\in \mathcal{B}$,
\begin{equation}
    z(\mathbf{K},\mathbf{X}) = \sum_{\mathbf{G}} \hat{z}_{\mathbf{G}} (\mathbf{K})e^{\iota(\mathbf{K}+\mathbf{G})}\cdot\mathbf{X},
    \label{eq:bloch1}
\end{equation}
and the sum is taken over all the vectors in the reciprocal lattice, i.e., for all integers $m$ and $n$, s.t., $\mathbf{G} = m \mathbf{G}_1 + n\mathbf{G}_2$.  Here $\iota^2=-1$.

Due to periodicity, it is sufficient to restrict ourselves to $\mathbf{X}\in \mathcal{W}$ \eqref{eq:defWg}. 
We plug \eqref{eq:bloch1} into \eqref{eq:flineqn} and, after multiplying the resulting equation by $e^{-\iota(\mathbf{K}+{\mathbf{G}})\cdot\mathbf{X}}$ (for a fixed reciprocal lattice vector ${\mathbf{G}}$), we then integrate 
over the Wigner-Seitz unit cell $\mathcal{W}$, to obtain:
\begin{equation}
    \ell^2  D(\mathbf{K},\mathbf{G})\hat{z}_{\mathbf{G}}(\mathbf{K}) + \sum_{\mathbf{G}'}\hat{z}_{\mathbf{G'}}(\mathbf{K})f(\mathbf{K},\mathbf{G}')=0,    
    \label{eq:simplify2}
\end{equation}
where $\ell^2$ is the area per unit cell of the lattice (highlighted in gray in Fig.~\eqref{fig:triangular lattice}),
\begin{equation}
D(\mathbf{K},\mathbf{G}):=\kappa |\mathbf{K}+\mathbf{G}|^4 + 2\mg{\gamma}|\mathbf{K}+\mathbf{G}|^2,
\end{equation} 
and 
\begin{equation}
    f(\mathbf{K},\mathbf{G}') = 2\sum_{j\neq 0}  \mg{A}_{ij}\Big[1 -\cos((\mathbf{K}+\mathbf{G}')\cdot\mg{\mathbf{X}}_j)\Big].
\end{equation}
Here $\mathbf{K}$ belongs to the first Brillouin zone $\mathcal{B}$, while $\mathbf{G}, \mathbf{G}'$ belong to the reciprocal lattice.
Details of the above calculation can be found in Supplementary Material.

Let us denote the multi-index $M=(m_1,m_2)$ and $\mathbf{G}_M = m_1\mathbf{G}_1 + m_2\mathbf{G}_2$ where $M\in \mathbb{Z}^2$. For numerical evaluation, we assume that $M$ belongs to a suitable finite grid; see the remark at the end of Sec.~\ref{sec:results}.
Using this notation, for any $\mathbf{K}\in \mathcal{B}$, let
\begin{equation}
    D_M := \ell^2 D(\mathbf{K},\mathbf{G}_M),\;f_M \equiv f(\mathbf{K},\mathbf{G}_M),\;\hat{z}_M = \hat{z}_{\mathbf{G}_M}(\mathbf{K}).
\end{equation}
Thus, we can re-write \eqref{eq:simplify2} as the matrix equation
\begin{equation}
\mathbf{L}\mathbf{z}:=
    \left(
    \begin{array}{cccc}
    D_1+f_1 & f_2 & f_3 & \cdots \\
    f_1 & D_2 + f_2 & f_3 & \cdots\\
    f_1 & f_2 & D_3 + f_3 & \cdots \\
    \vdots & \vdots & \vdots & \ddots
    \end{array}
    \right)\left(
    \begin{array}{c}
    \hat{z}_1\\\hat{z}_2\\\hat{z}_3\\\vdots
    \end{array}
    \right)=\mathbf{0},
    \label{eq:simplify3}
\end{equation}
where the subscripts $1,2,\cdots$ denote a natural ordering of the tuple $M$. Nontrivial solutions of \eqref{eq:simplify3} exist when the determinant of the matrix in the equation is zero. To facilitate numerical computation, we normalize $\mathbf{L}$ in the results produced via numerical evaluation by dividing row $i$ by $D_i$.

To compute the determinant, we express the matrix $\mathbf{L}$ as
\begin{equation}
\mathbf{L} = \mathbf{D}+\mathbf{F}, 
\label{defL}
\end{equation}
where $\mathbf{D}$ is the diagonal matrix
\begin{equation}
\mathbf{D} = \left(
    \begin{array}{ccc}
    D_1 & 0 & \cdots\\
    0 & D_2 & \cdots\\
    \vdots & \vdots & \ddots
    \end{array} 
    \right),
\end{equation}
and 
\begin{equation}
    \mathbf{F} = \left(
    \begin{array}{cccc}
    f_1 & f_2 & f_3 & \cdots\\
    f_1 & f_2 &  f_3 & \cdots\\
    \vdots & \vdots & \vdots & \ddots
    \end{array} 
    \right) = \left(
    \begin{array}{c}
    1\\1\\ 1 \\ \vdots
    \end{array} 
    \right)\left(
    \begin{array}{cccc}
    f_1 & f_2 & f_3 & \cdots
    \end{array} 
    \right).
\end{equation}
In this way, we express $\mathbf{L}$ as a rank-one update to $\mathbf{D}$.
Using a well-known identity (\emph{matrix-determinant lemma}), we obtain
\begin{equation}
\det(\mathbf{L}) = \left(1 + \sum_M \frac{f_M}{D_M}\right) \Pi_{M} D_M.
\label{eqdetL}
\end{equation}
We formally interpret the determinant as a functional determinant of the operator $\mathbf{L}$. Of course, for numerical evaluation in Sec.~\ref{sec:results}, we approximate it as a classical determinant, as the multi-index $M$ belongs to a finite grid.

\begin{remark}
Note that the case $\mg\gamma<0$ is associated with a compressive stress in the plane of the membrane. This could lead to buckling instabilities that lead to out-of-plane displacements. Here, we are primarily concerned with how the particles can induce out-of-plane displacement in the membrane. For this reason, we set $\mg\gamma\geq 0$. It then follows that $D_M\geq 0$. Therefore, according to \eqref{eqdetL}, the zeros of $\det(\mathbf{L})$ can be computed by setting
\begin{equation}
    C(\kappa,\mg\gamma,\mathbf{K}):=1 + \sum_M \frac{f_M}{D_M} = 0.
    \label{eq:def C}
\end{equation}
\end{remark}

\section{Results}
\label{sec:results}
In this section, we show that the trivial flat state loses stability, via Bloch modes, at certain critical values of $\kappa$ and $\mg{\gamma}$. In particular, the instabilities at critical $\kappa$ and $\mg{\gamma}$ are associated with the birth of new periodic non-planar states.  We compute the stability boundary of the trivial state with respect to these new configurations that bifurcate from it. 

In light of the remark at the end of Sec.~\ref{ssec:linearization}, since $\tau=0$, up to first order in $\xi$ (the amplitude of perturbation), we find that $\gamma=\mg{\gamma}$. Therefore, taking this into account in what follows, we suppress the decoration on $\gamma$ and simply write $\gamma$ instead of $\mg{\gamma}$.

Towards the goal of finding the stability boundary of the trivial state, with respect to some new configurations that bifurcate from it,
let us first note that for large values of $\kappa$ or $\gamma$, all the eigenvalues of $\mathbf{L}$ \eqref{defL} are positive. This is because, as $\kappa\to\infty$ (or ${\gamma}\to \infty$), $\mathbf{L}$ approaches a diagonally dominant matrix with positive diagonal entries so that all the eigenvalues of $\mathbf{L}$ are positive. Alternatively, this can also be inferred by noting that as $\kappa\to\infty$, the dominant part of Euler-Lagrange equation \eqref{eq:flineqn} is biharmonic operator, whose eigenvalues are also positive. Similarly, if $\gamma\to \infty$, the dominant part of Euler-Lagrange equations is the harmonic operator (the negative of Laplacian operator), whose eigenvalues are all positive for given periodic boundary conditions.  Thus, for large values of $\kappa$ or $\gamma$, the trivial solution is stable (to periodic perturbations). Physically, this can be understood by observing that if the bending stiffness or the surface tension is high, it is difficult to deform the membrane from its flat state. 

Thus, for large values of $\kappa$ or $\gamma$, the trivial flat configuration is a (locally) \emph{stable} equilibrium. We have inferred the stability of the trivial state from the fact that for this state, the linearization \emph{is the second variation}, and it has positive eigenvalues, implying local stability.  Note that our Bloch ansatz \eqref{eq:bloch1} only includes periodic perturbations, and therefore, when we use the term \emph{stability}, we mean local stability to periodic perturbations. 

As $\kappa$ or $\gamma$ is decreased, we verify numerically that one of the eigenvalues of $\mathbf{L}$ passes through the zero and becomes negative, signaling an onset of instability.  Equivalently, we detect when the determinant of $\mathbf{L}$, or its proxy $C$, cf. \eqref{eq:def C}, changes sign. Observe that $C$ ($\equiv C(\mathbf{K})$ for fixed $\kappa,\gamma$ \eqref{eq:def C}) is a symmetric function of $\mathbf{K}$ and has the symmetry of the reciprocal lattice. Thus, it suffices to evaluate $C$ on a unit cell which is a fundamental unit cell. Numerically, we find that the instability modes always arise on the boundary of the unit cell, and are in fact, on the $K_x$ axis (where $K_x$ is the $x$-component of the wave vector $\mathbf{K}$, i.e., $\mathbf{K}=K_x\mb{e}_1+K_y\mb{e}_2\in\mathbb{R}^2$). 

To determine the threshold of instability in the $\kappa-\gamma$ plane, we detect the values of those parameters at which $C$ changes sign (i.e., attains a value of zero). Recall that if $\kappa$ and $\gamma$ are large, then $C(K_x\mb{e}_1)$ is positive for all $K_x$ and as these parameters are decreased, $C$ changes sign at some critical value of $K_x$. This necessarily happens at the minimum value of $C$.  We detect this change of sign as follows.  First, we fix one of the parameters, say, $\kappa$, and minimize $C$ as a function of  $\gamma$ and $K_x$. Then, to determine when the minimum value of $C$ crosses zero, we perform a 1D search using the bisection method, starting with a large interval of search for $\gamma$. In this manner, for a given value of $\ell$, by changing the values of $\kappa$, we determine the value of $\gamma$ for which the trivial state loses stability via certain Bloch mode. 

\begin{figure}[h!]
    \centering
    \includegraphics[width=.8\textwidth]{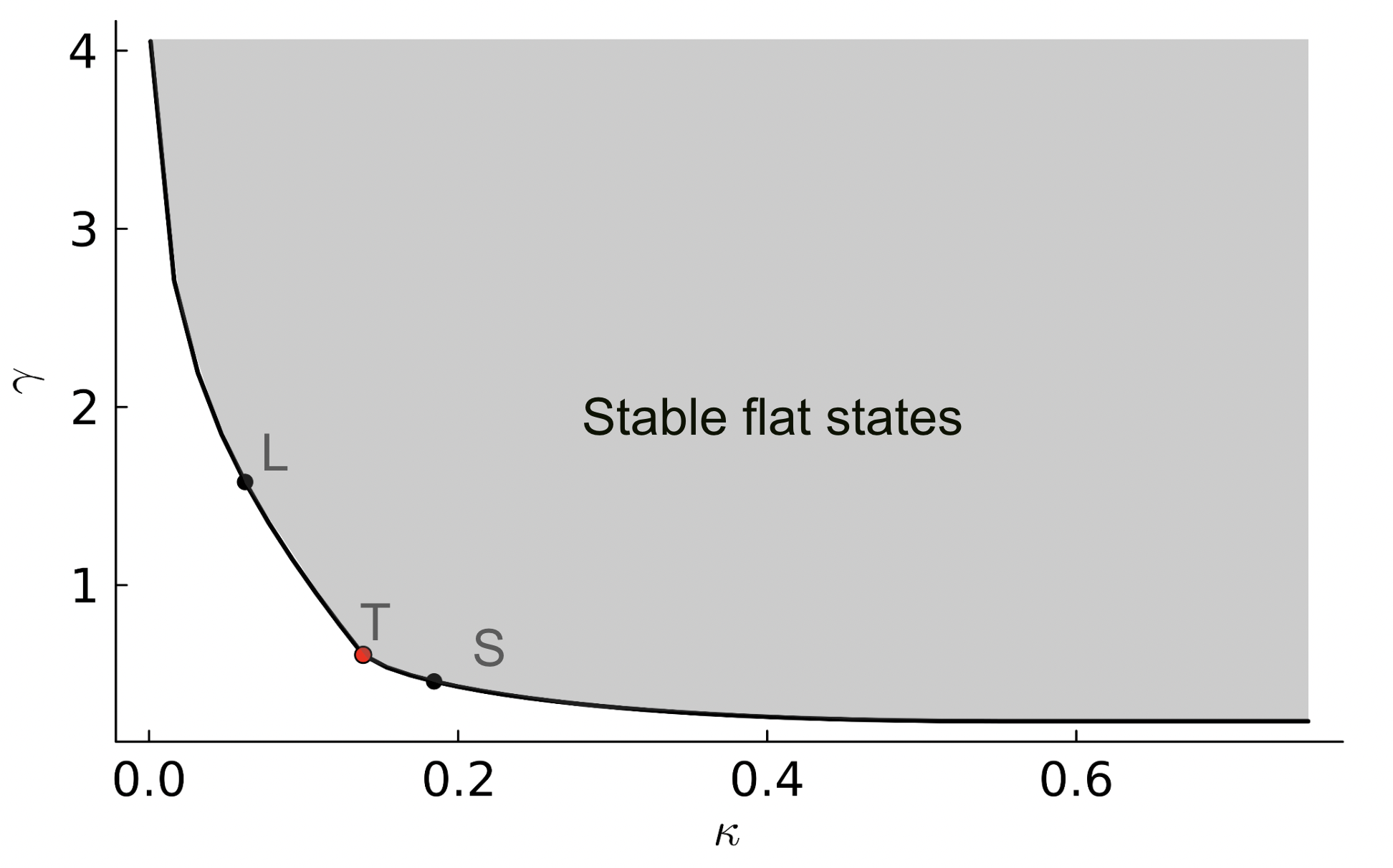}
    \caption{Curve of instability in the $\kappa-\gamma$ plane for $\ell=0.92$. The point $T$ marks the transition between large wavenumber modes ($L$) to small wavenumber modes $(S)$.}
    \label{fig:generic-kappa-gamma}
\end{figure}

\begin{figure*}[h!]
    \begin{tabular}{ccc}
         \includegraphics[width=.3\textwidth]{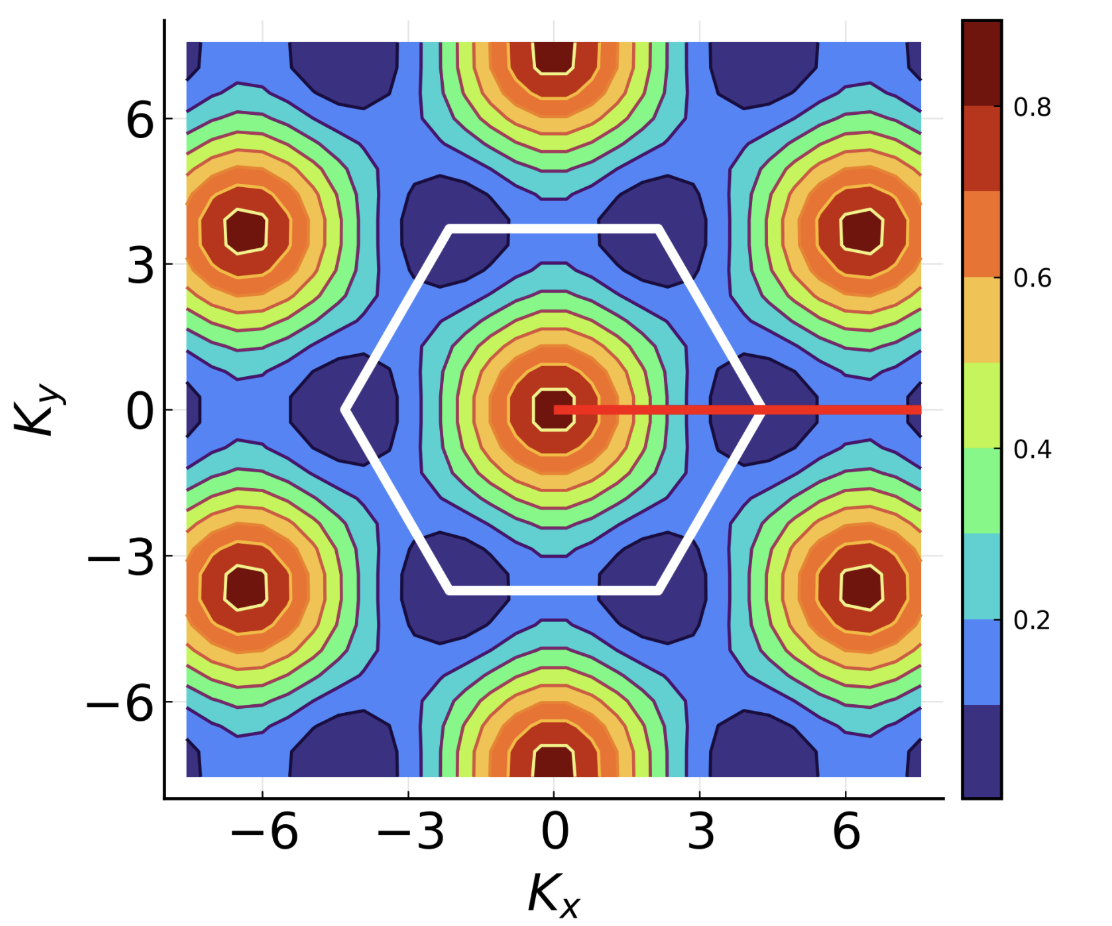} &
         \includegraphics[width=.33\textwidth]{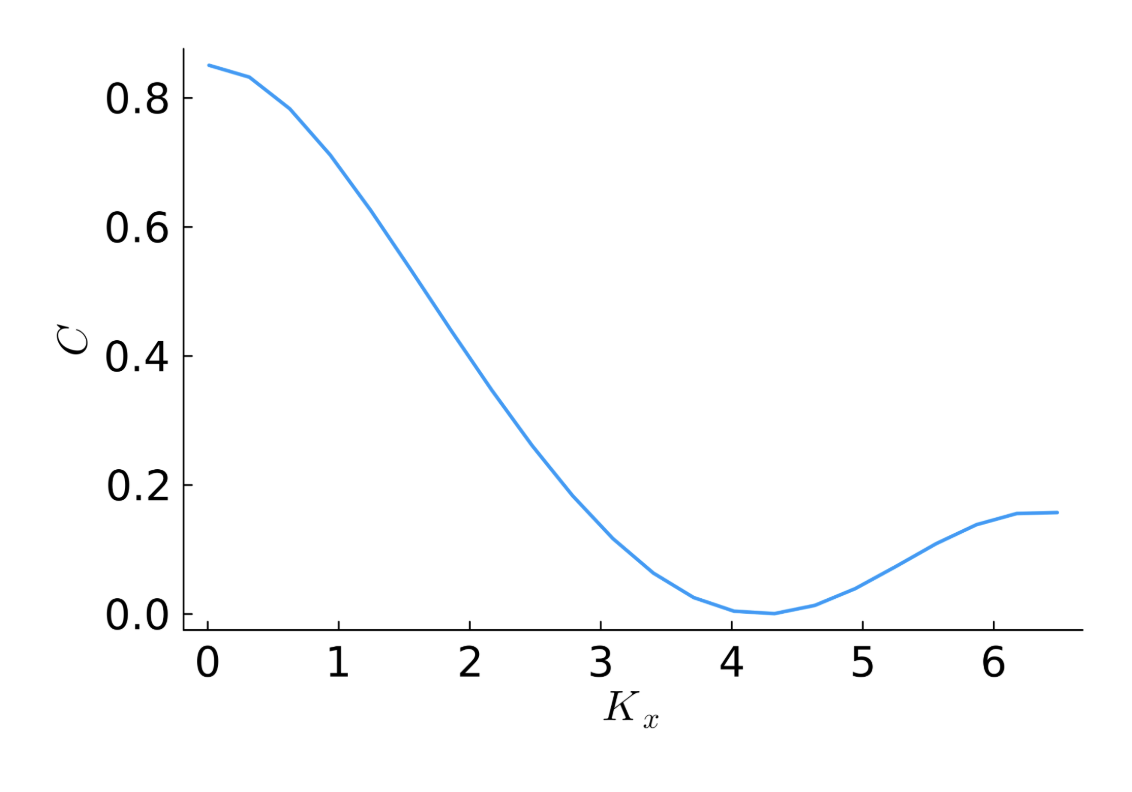} &
          \includegraphics[width=.3\textwidth]{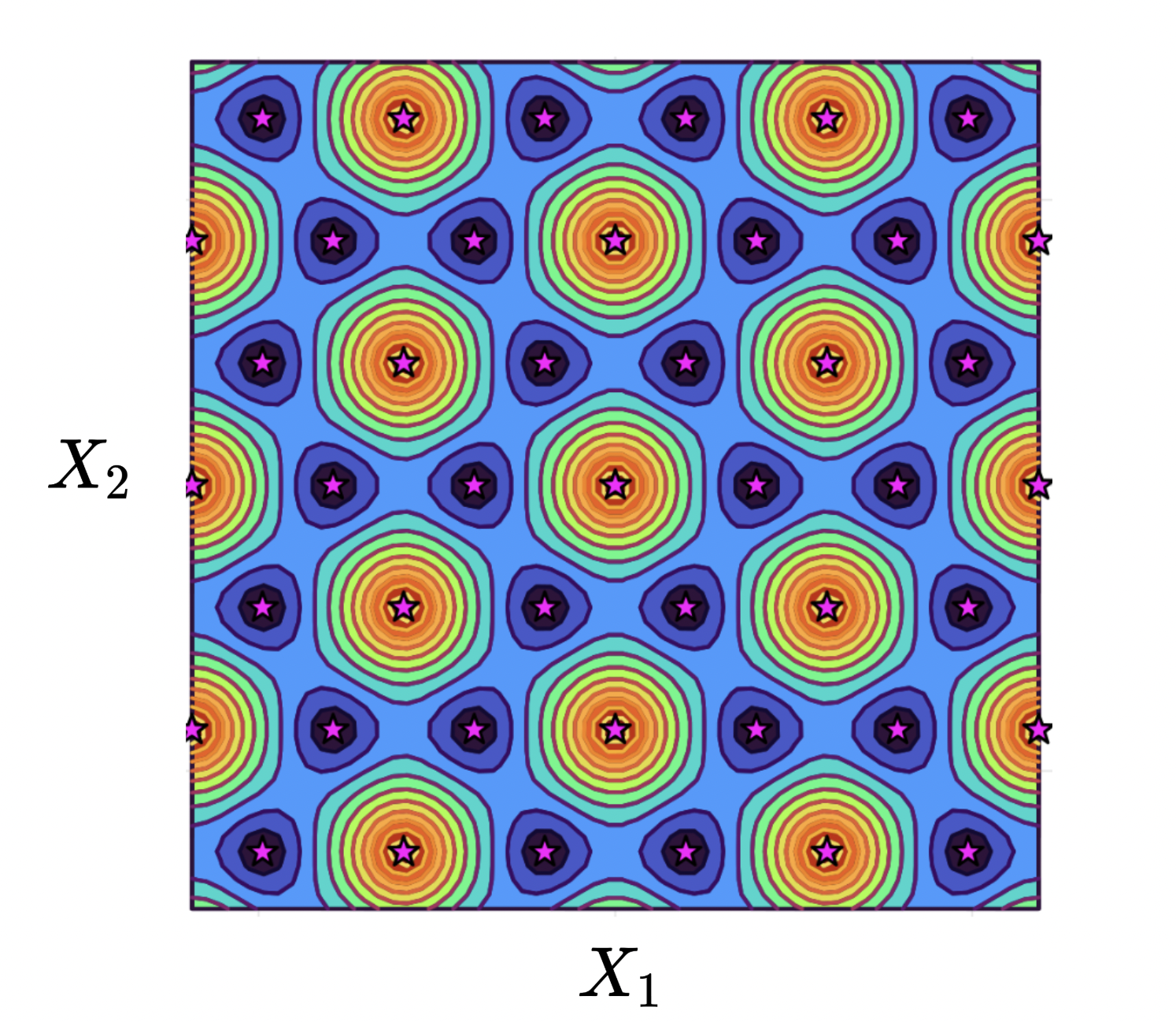} \\
         (a) & (b) & (c) 
    \end{tabular}
    \caption{Plots corresponding to the $L$ mode in Fig.~\ref{fig:generic-kappa-gamma}. (a) Contour plots of $C$ versus $\mathbf{K}$. (b) Plot of $C$ along $K_x$ axis showing a zero at $K_x\approx 4.5$. (c) Contour plot of $z(X_1,X_2)$ of the bifurcating state (null vector).}
    \label{fig:L panel}
\end{figure*}

The stability threshold in the $\kappa-\gamma$ plane, obtained using the aforementioned procedure, generically looks like the curve shown in Fig.~\ref{fig:generic-kappa-gamma} (for which we have used $\ell=0.92$). For parameter values above this curve in the $\kappa-\gamma$ plane (shown in gray in the figure), the trivial solution is a stable equilibrium (all eigenvalues of the $\mathbf{L}$ are positive), and below the curve (in the region towards $(0,0)$), the trivial state is unstable. At the threshold, we find that one of the eigenvalues is zero, thus signaling the birth of a new state; a bifurcation occurs. Since the curve separates distinct configurations, we shall call this a \emph{transition curve}. Every $(\kappa,\gamma)$ on the transition curve denotes parameter values for which a bifurcation occurs. The bifurcated state can be characterized by the null vector of $\mathbf{L}$ and has a characteristic wavenumber $\mathbf{K}$ that is the minimizer of $C(\mathbf{K})$. As we shall see below, the new state corresponds to non-planar modes of the membrane. 

\begin{figure*}[h!]
    \begin{tabular}{ccc}
         \includegraphics[width=.3\textwidth]{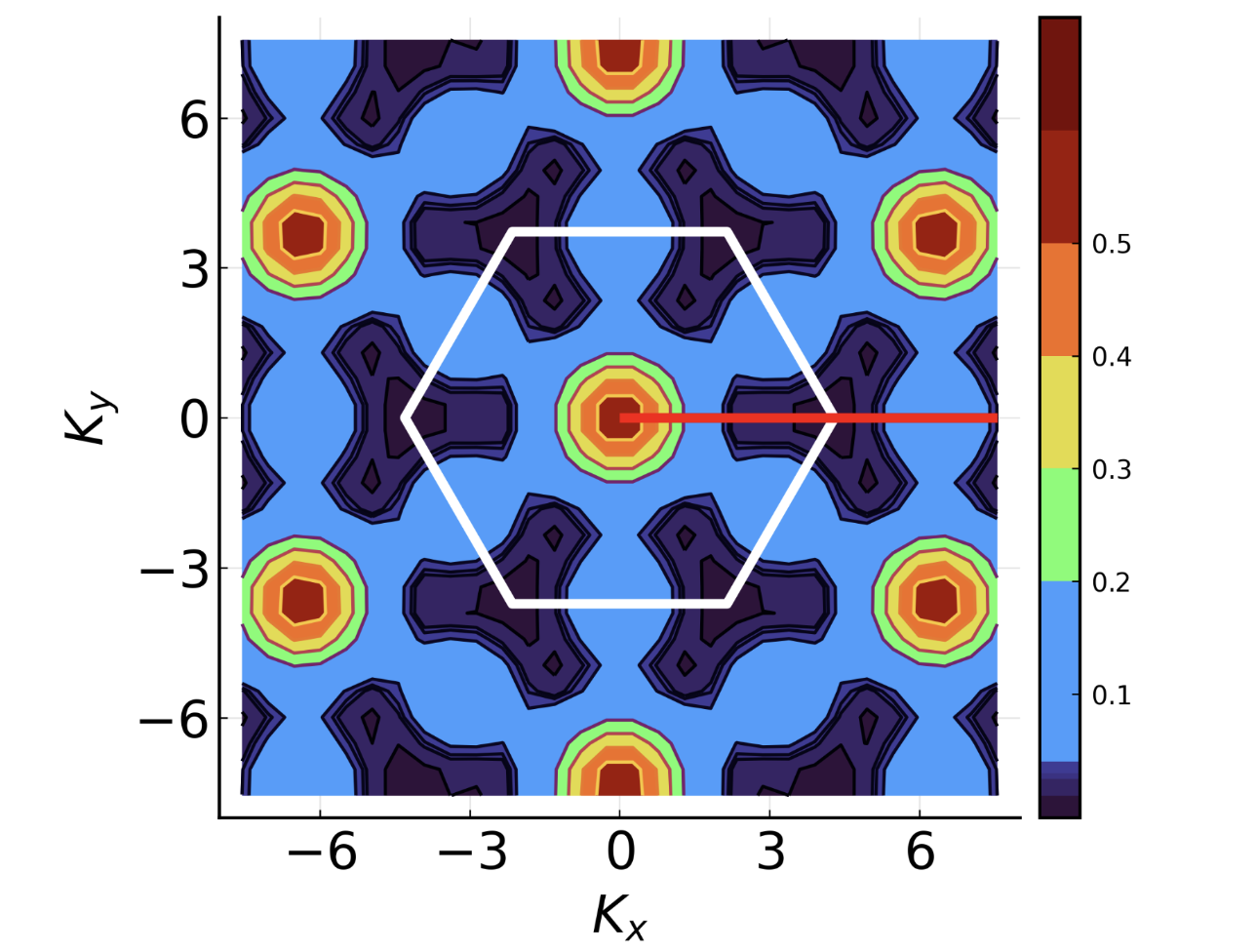} &
         \includegraphics[width=.33\textwidth]{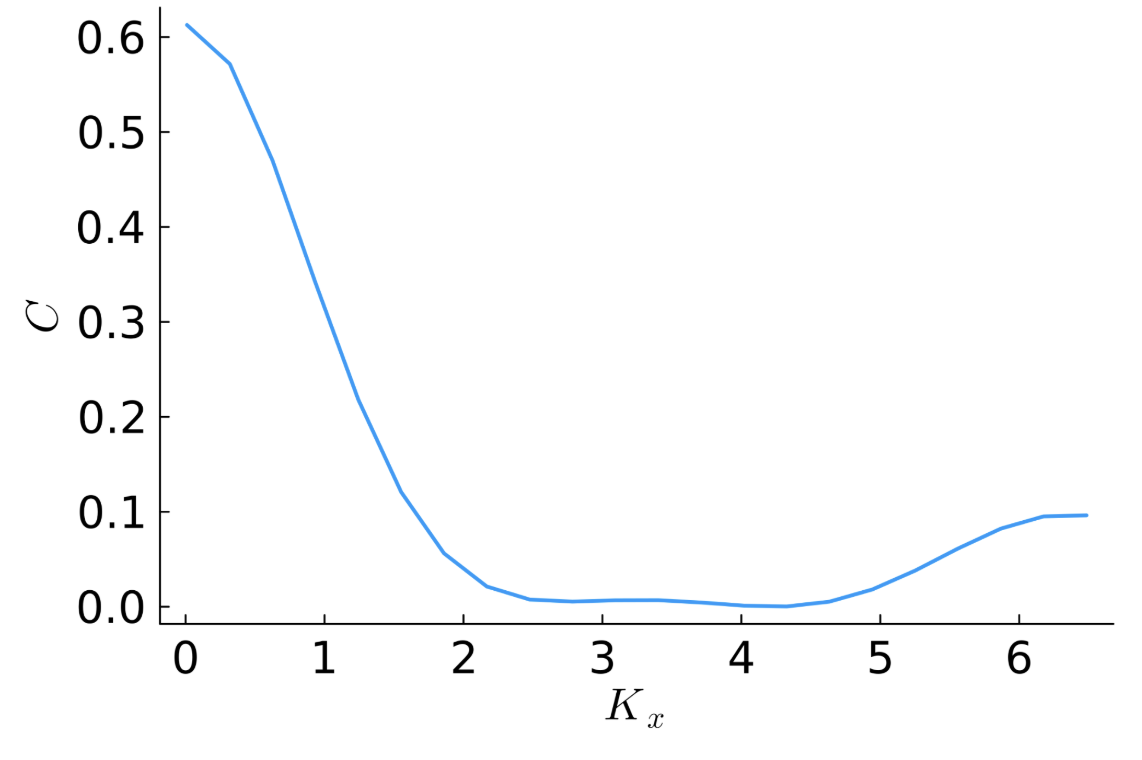} &
          \includegraphics[width=.3\textwidth]{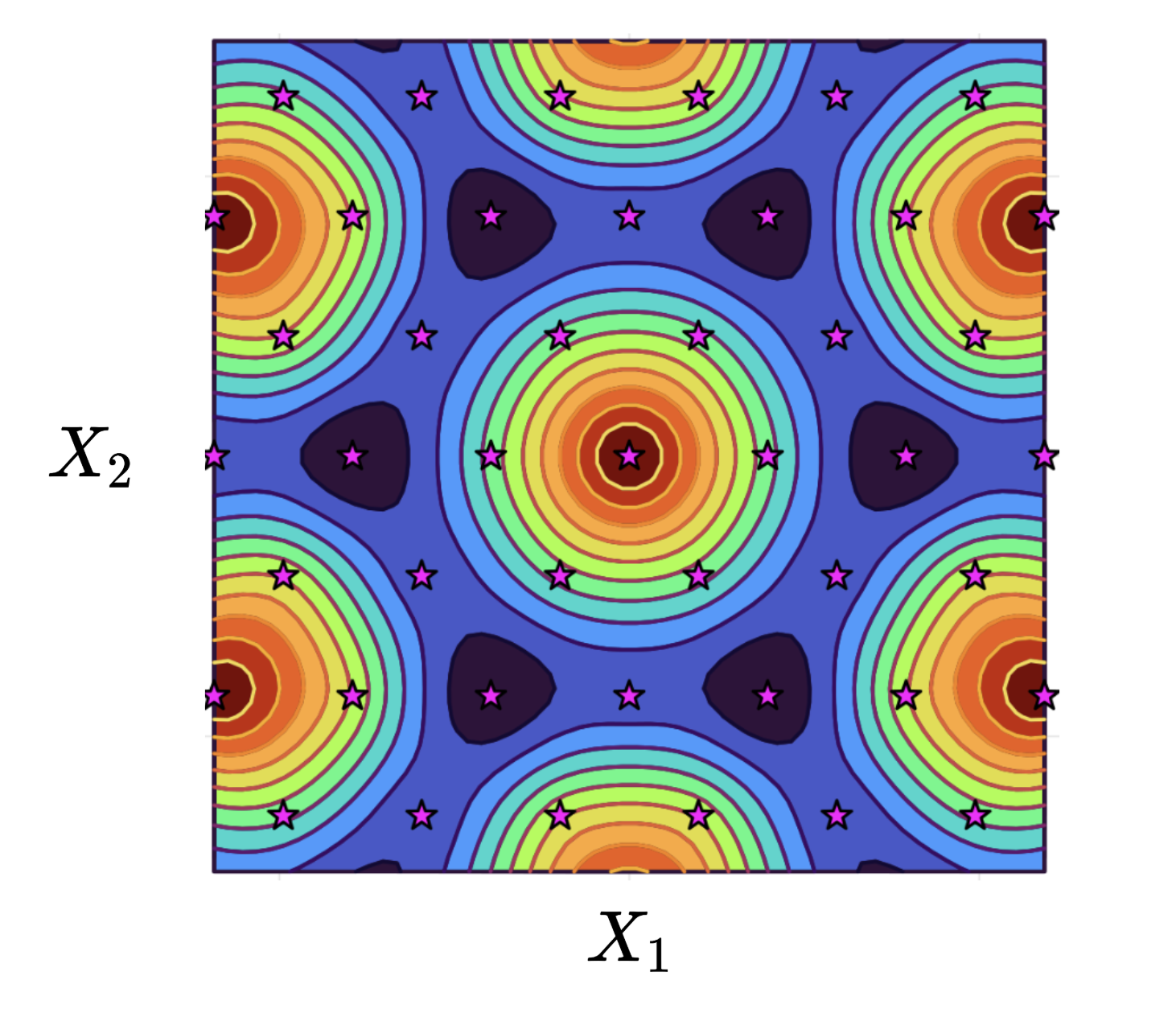} \\
         (a) & (b) & (c) 
    \end{tabular}
    \caption{Plots corresponding to the point $T$ in Fig.~\ref{fig:generic-kappa-gamma}. (a) Contour plots of $C$ versus $\mathbf{K}$. (b) Plot of $C$ along $K_x$ axis showing two zeros at $K_x\approx 2.5$ and $4.5$. (c) Contour plot of $z(X_1,X_2)$ of the bifurcating state (null vector) showing a bud-like formation.}
    \label{fig:T panel}
\end{figure*}

We observe that the transition curve has a discontinuous slope at a point labeled $T$ in Fig.~\ref{fig:generic-kappa-gamma} (red colored point). This point marks the transition from bifurcated states with large wavenumber modes to modes with small wavenumber. As can be seen in Fig.~\ref{fig:generic-kappa-gamma}, large wavenumber modes occur for smaller values of $\kappa$ (or large values of $\gamma$) and small wavenumber modes occur for larger values of $\kappa$ (or small values of $\gamma$). 

\begin{figure*}[h!]
    \begin{tabular}{ccc}
         \includegraphics[width=.3\textwidth]{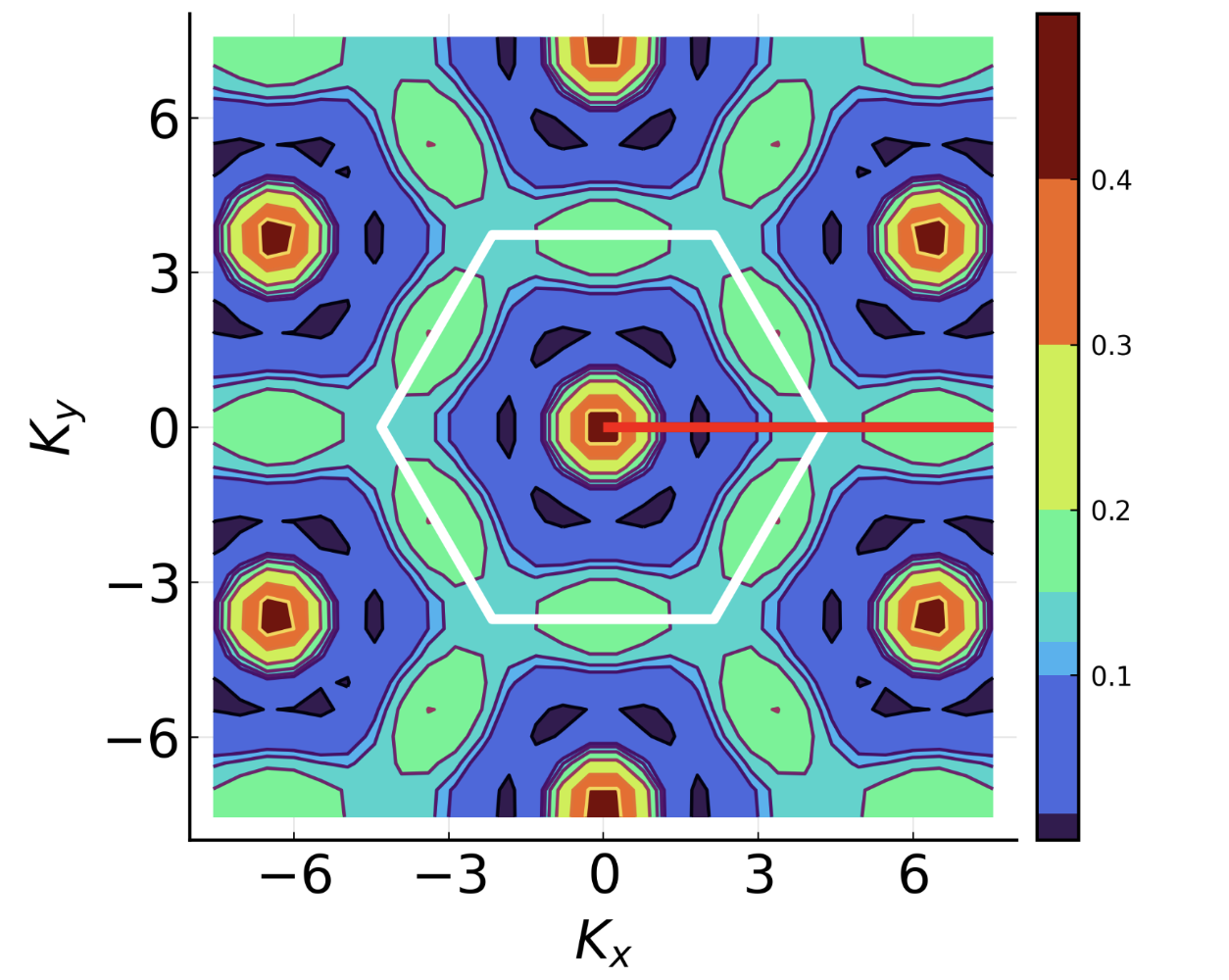} &
         \includegraphics[width=.33\textwidth]{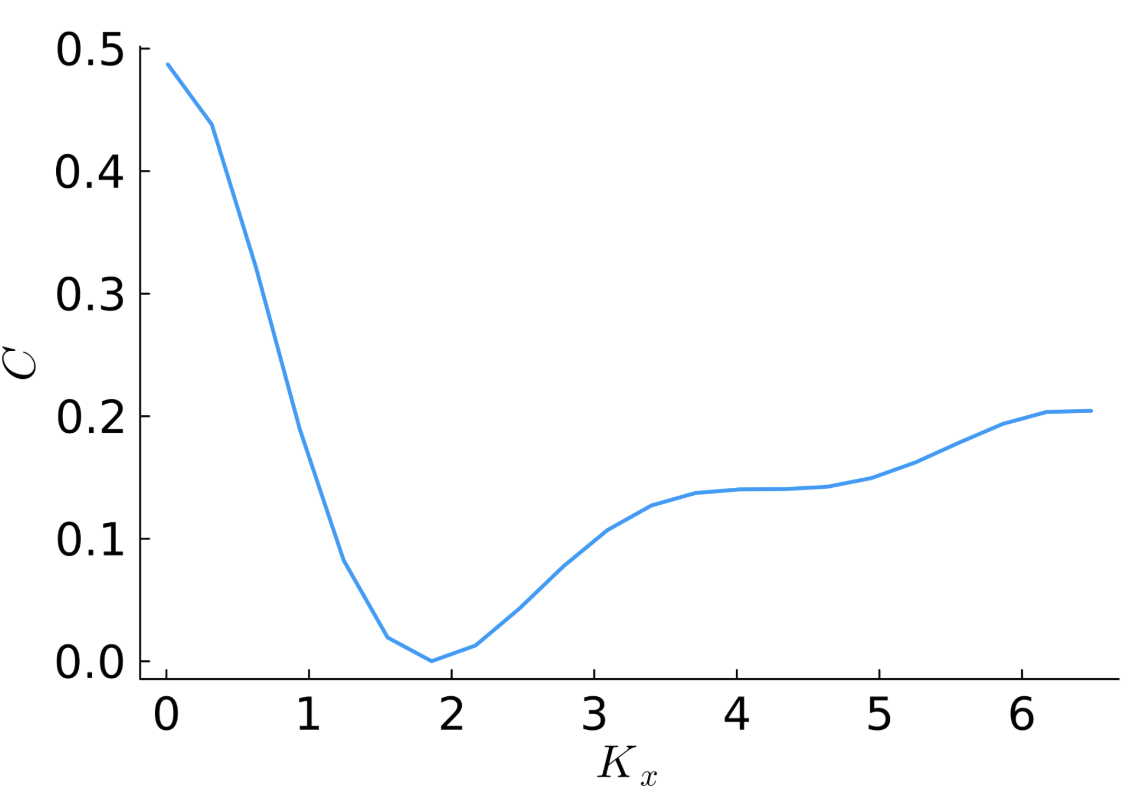} &
          \includegraphics[width=.3\textwidth]{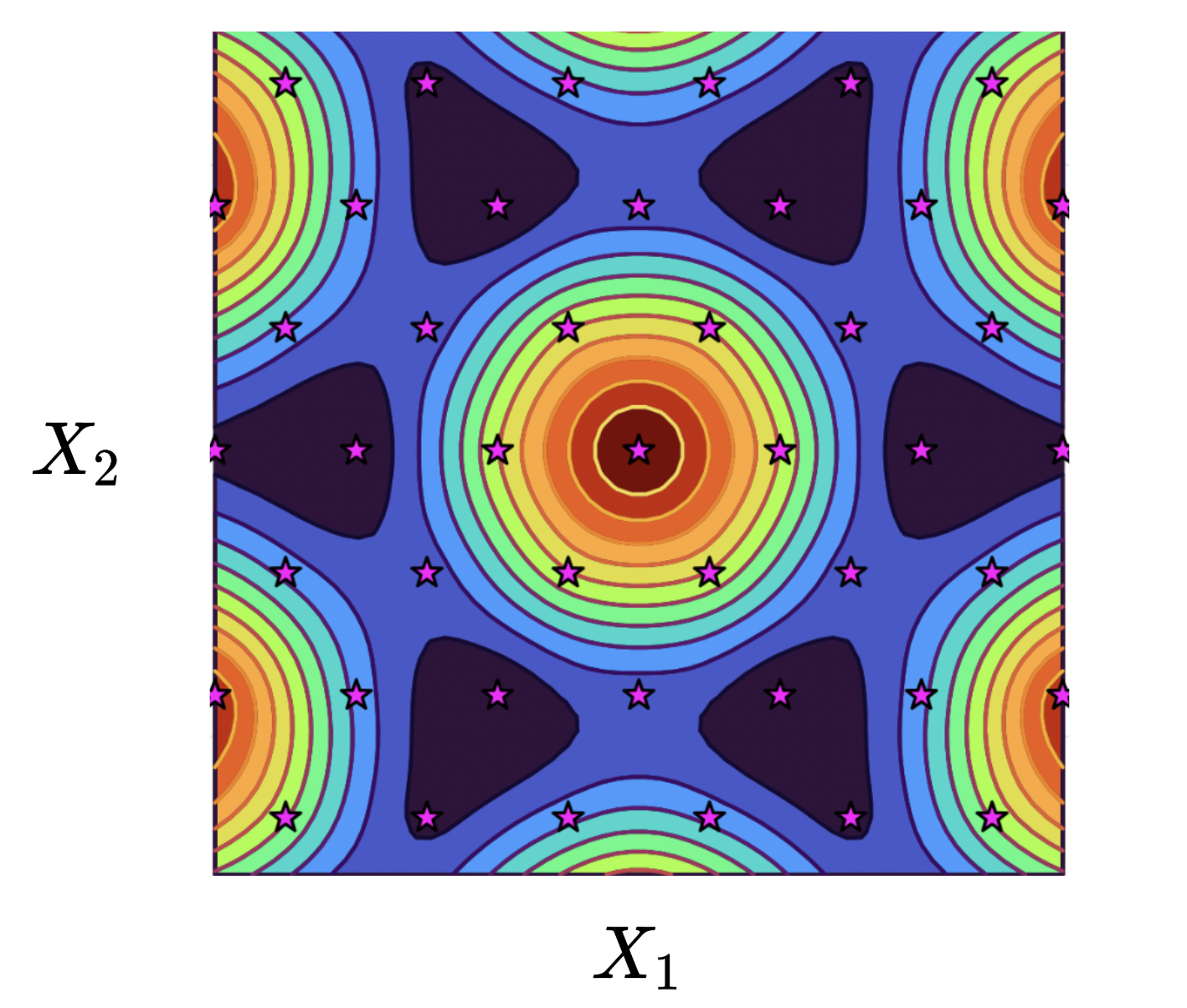} \\
         (a) & (b) & (c) 
    \end{tabular}
    \caption{Plots corresponding to the $L$ mode in Fig.~\ref{fig:generic-kappa-gamma}. (a) Contour plots of $C$ versus $\mathbf{K}$. (b) Plot of $C$ along $K_x$ axis showing a zero at $K_x\approx 2$. (c) Contour plot of $z(X_1,X_2)$ of the bifurcating state (null vector) showing a bud-like formation.}
    \label{fig:S panel}
\end{figure*}

To better understand the transition from $L$ to $S$, let us consider Figs.~\ref{fig:L panel}, \ref{fig:T panel}, and \ref{fig:S panel} that correspond to points $L$, $T$, and $S$, respectively, of Fig.~\ref{fig:generic-kappa-gamma}. Each of the three figures has three subplots, labeled (a), (b), and (c). In (a), we plot the contour map $C$ as a function of $\mathbf{K}$, in (b), we plot $C$ along the $K_x$ axis, and in (c), we plot the null vector that corresponds to the zero of $C$. Let us first consider Fig.~\ref{fig:L panel} which represents the large wavenumber mode. In (a), we see that $C$ is non-negative everywhere and attains a minimum value of zero (shown in dark blue) at the six corners of the Brillouin zone (shown as a white hexagon). In the corresponding panel (b) of Fig.~\ref{fig:L panel}, we plot $C(K_x\mb{e}_1)$ along the $K_x$ axis). Note that the wavenumber corresponding to the zero eigenvalue is around $K_x\approx 4.5$. The contour plot of the eigenmode $z(X_1,X_2)$ is plotted in panel (c), which we obtained by computing the null vector of $\mathbf{L}$ and plugging it into the Bloch ansatz \eqref{eq:integral z} and \eqref{eq:bloch1}. The blue regions correspond to membrane deformation into the plane, while the orange regions correspond to out-of-plane deformation. Note that due to the inherent reflection symmetry about the $X_1-X_2$ plane in this problem, the mirror reflection of Fig.~\ref{fig:L panel}(c) is also a solution. Thus, we expect the bifurcation to be a pitchfork bifurcation. The exact numerical values of the color bar are not relevant, as this is an eigenmode. The particles, plotted as `stars', form a triangular lattice. Observe that for each particle that pops out of the plane, its six nearest neighbors pop into the plane, giving the surface a corrugated `egg-carton-like' appearance. We term such modes \emph{$L$-modes}. As we vary $\kappa$ and $\gamma$ along the transition curve of Fig.~\ref{fig:generic-kappa-gamma}, curiously, we find the bifurcating states on the $L$ branch always have a wavenumber close to $K_x\approx 4.5$. That is, the contour plot of the null vectors generically appears as Fig.~\ref{fig:L panel}(c).

As $\kappa$ is increased (and $\gamma$ is decreased along the transition curve), we reach the transition point $T$ (in Fig.~\ref{fig:generic-kappa-gamma}). The minimum of $C$ that was previously (for $L$ states) located at the corner of the first Brillouin zone splits into three minima (only one of which lies within the Brillouin zone). This is shown Fig.~\ref{fig:T panel}(a) where we see the blue triangular region containing the minima near the six vertices of the Brillouin zone. The plot of $C$ versus $K_x$ in Fig.~\ref{fig:T panel} shows a 1D view of the minimum splitting. The single minimum of the $L$ state splits into two minima located at $K_x\approx 2.5$ and $K_x\approx 4.5$. The eigenmode that corresponds to the bifurcated state with a linear combination of $K_x\approx 2.5$ and $4.5$ is shown in Fig.~\ref{fig:T panel}(c). We observe that, unlike in the large wavenumber case, for the $T$ state, the particles are no longer confined to the maxima and minima of $z$. Instead, we see a bud-like protrusion on the membrane surface consisting of a cluster of six particles. The periodic pattern repeats beyond the unit cell.

Further increasing $\kappa$ along the transition curve, we move into the $S$ branch. The zeros of the $C$ start moving into the first Brillouin zone, but remain on the line connecting the origin to the vertex of the Brillouin zone. This can be seen in Fig.~\ref{fig:S panel}(a). The 1D section of the graph of $C$ along $K_x$ axis is shown in Fig.~\ref{fig:S panel}(b). Note the minimum at $K_x\approx 2$ with $C\approx 0$. Also, observe that the other minimum located at $K_x\approx 4.75$ is no longer a root of $C$. Thus, there is only one null vector with a wavenumber $K_x\approx 2$. The eigenmode $z(X_1,X_2)$ is plotted in Fig.~\ref{fig:S panel}(c). Note that the size of the bud has grown compared to the $T$ state. We will call such bud-shaped modes \emph{$S$-modes}. In contrast to the $L$ case, increasing $\kappa$ (along the $S$ branch of Fig.~\ref{fig:generic-kappa-gamma}) will shift the minimum to lower values of $K_x$, hence the term `short wavenumber mode'. That is, moving to large $\kappa$ values will result in large, bud-like caps on the membrane surface.

\begin{figure}[h!]
    \centering
    \includegraphics[width=.8\textwidth]{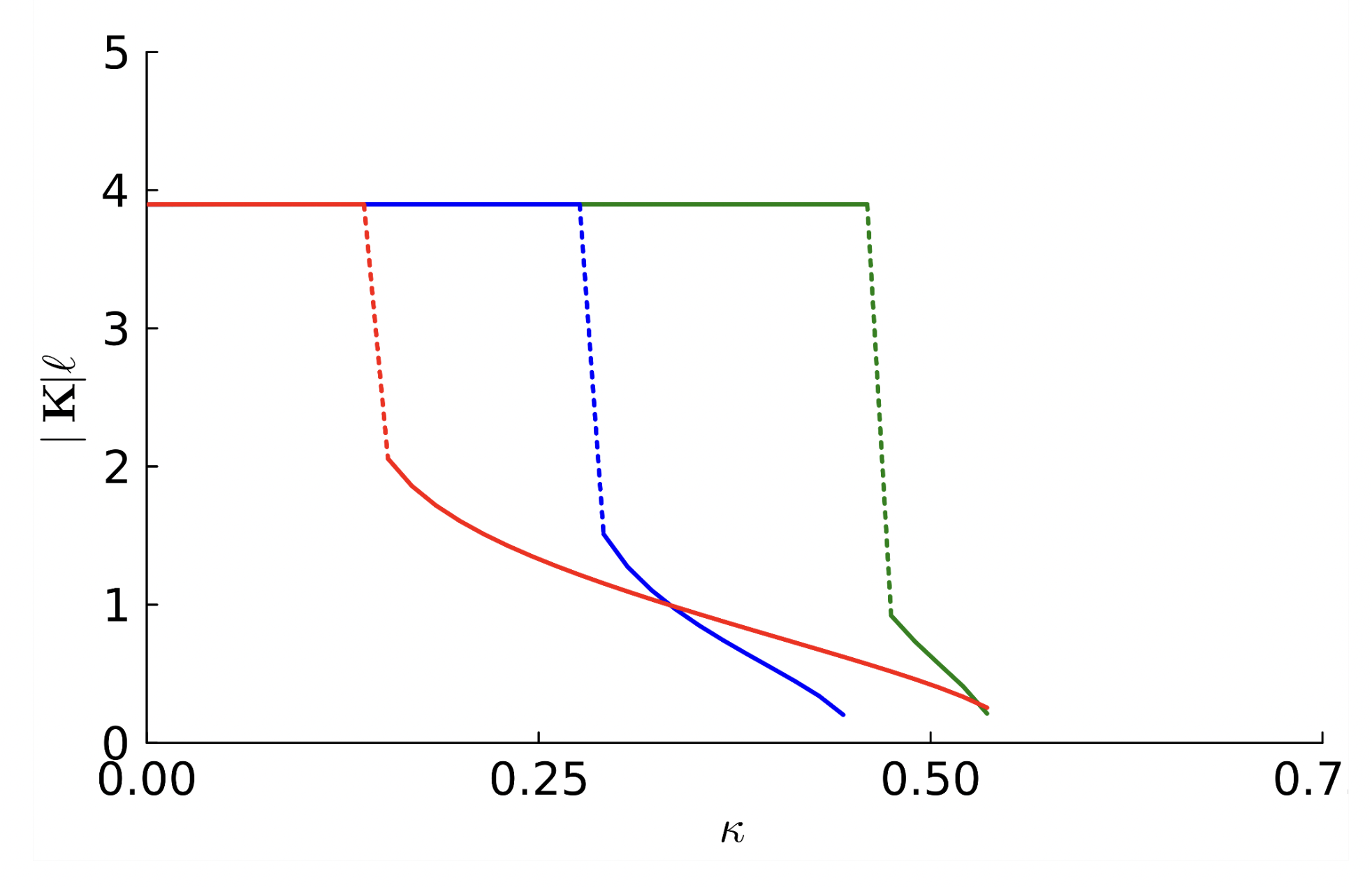}
    \caption{Wave number of the bifurcated eigen state plotted as a function of $\kappa$. Transition point $T$ of Fig.~\ref{fig:generic-kappa-gamma} is shown as a dashed line. The curves shown in red, blue, and green correspond to $\ell=0.92,0.905, 0.89$, respectively.}
    \label{fig:bifurcation curve}
\end{figure}

The previous observations can also be visualized by plotting the wavenumber of the bifurcating state (normalized by $\ell$) as a function of $\kappa$ (while simultaneously changing $\gamma$ along the transition curve). This is shown in  Fig.~\ref{fig:bifurcation curve}. The red curve corresponds to $\ell=0.92$, blue to $\ell=0.905$, and green to $\ell=0.89$. Each of these curves resembles a depiction of a first-order transition. Focusing on one of these curves, say the blue one, we see that for small values of $\kappa$, the wavenumber is a constant. This observation was already noted above. The transition point $T$ in Fig.~\ref{fig:generic-kappa-gamma} corresponds to the dotted line; we see an abrupt transition to a new branch with states having smaller wavenumbers, whose wavenumber decreases as $\kappa$ is increased. That is, the size of the bud seen in Fig~\ref{fig:S panel}(c) grows as $\kappa$ is increased.

\begin{figure}[h!]
    \centering
    \includegraphics[width=.8\textwidth]{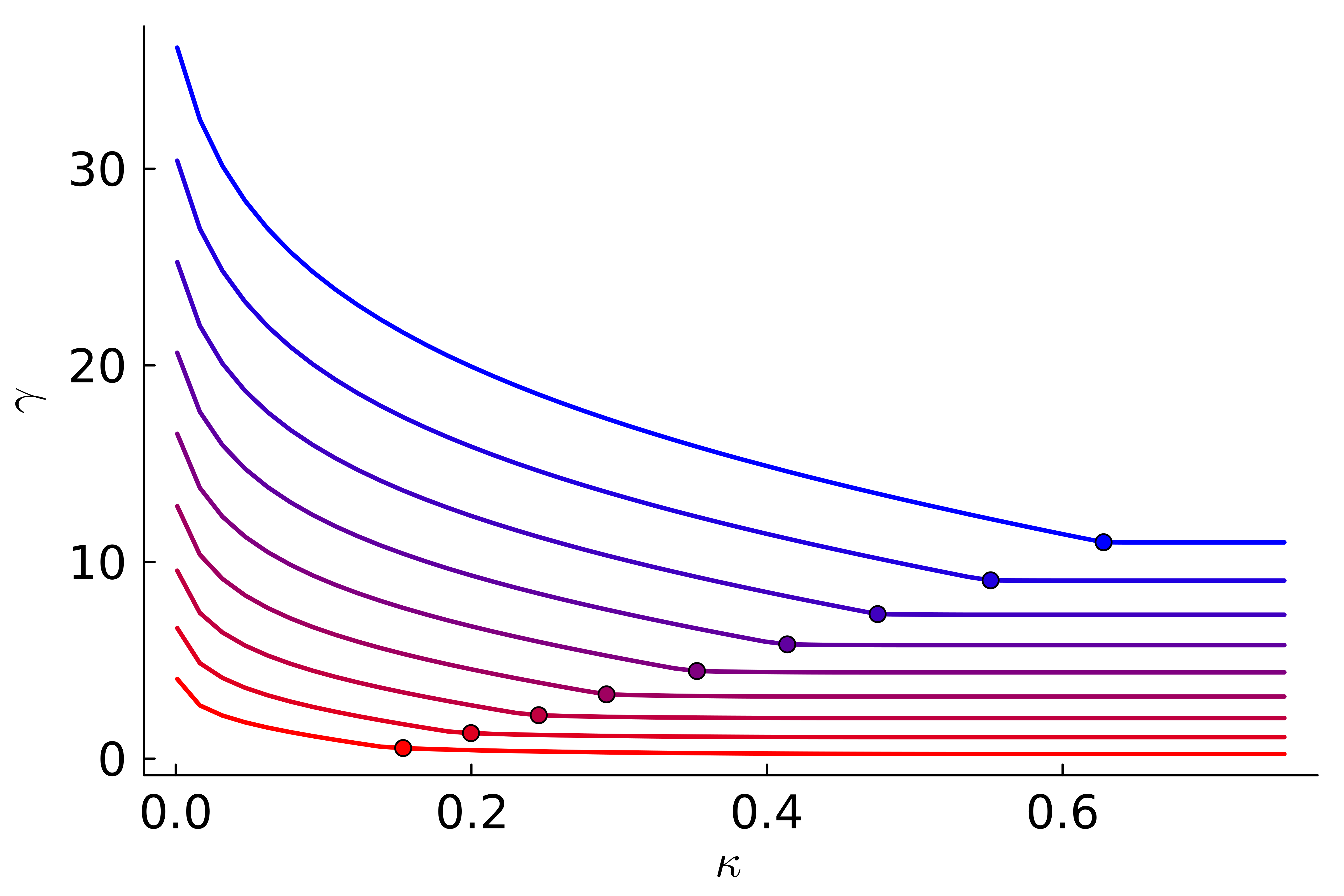}
    \caption{Dependence of the transition curve on $\ell$. Blue denotes smaller values of $\ell$ and red denotes larger values. Here $\ell$ lies in the range $[0.88,0.93]$.}
    \label{fig:effect of ell}
\end{figure}

The effect of $\ell$ on the stability curve is shown in Fig.~\ref{fig:effect of ell}. We plot this figure for $0.88\leq \ell\leq 0.93$. We see that as $\ell$ increases, the curve shifts downwards. This is also reflected in Fig.~\ref{fig:bifurcation curve} where we see that as $\ell$ increases, the transition state (shown as dashed lines) shifts to smaller $\kappa$ values. However, beyond a certain value of $\ell$, the transition curve disappears and is no longer in the $\gamma>0$ regime considered here. This can be understood as follows. We find that instability occurs only when $0.83\leq \ell\leq 0.93$, when the particles are under compression (i.e., repulsive regime of the LJ potential). The lattice of particles relieves this compressive stress by buckling out of the plane (either as an $L$-mode or an $S$-mode). We find that the buckling instability is not observed when the particles are under tension (for large values of $\ell$).

The transition curve's trend with respect to $\ell$, noted in the previous paragraph, can be understood as a competition between membrane bending and particle repulsion. If we decrease $\ell$, then the particles would be further compressed (since $r_e$ is fixed to be one). For small values of $\kappa$ (or large $\gamma$), the compressive stress could be relieved by buckling out of the plane in an $L$-mode (e.g., shown in Fig.~\ref{fig:L panel}(c)). However, for larger values of $\kappa$ (or small $\gamma$), large curvatures will be energetically expensive and the system prefers to relieve the compressive stress by buckling in a bud-shaped $S$-mode with smaller curvatures, Fig.~\ref{fig:S panel}.  It is interesting to note in Fig.~\ref{fig:bifurcation curve} that for different values of $\ell$, the $L$-modes, which correspond to the horizontal part of the curve, coincide when plotted with the normalized wavenumber.

\begin{figure}[h!]
    \centering
    \includegraphics[width=.8\textwidth]{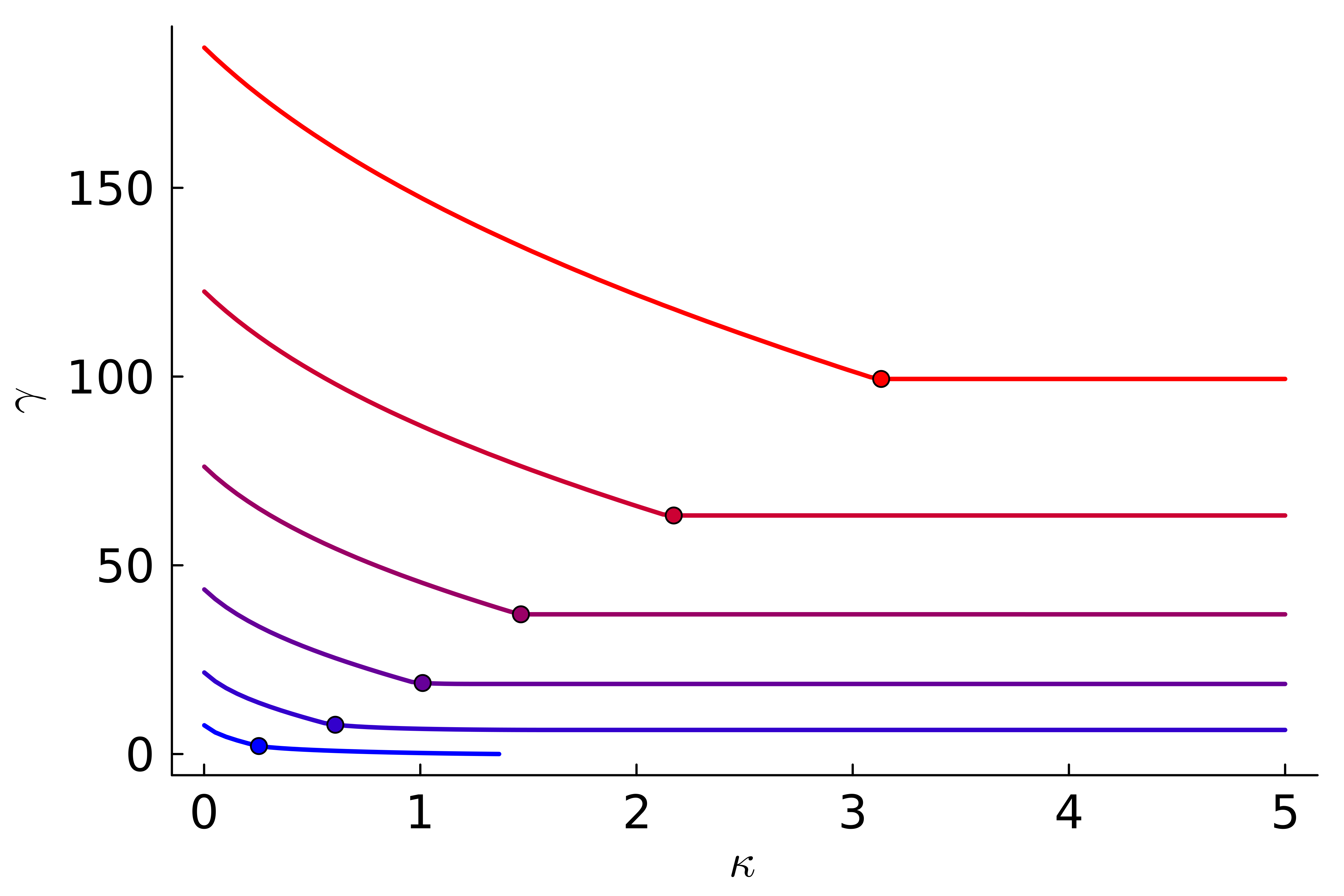}
    \caption{Dependence on the generalized-LJ parameter, $n$, we have set $\ell=0.62$}
    \label{fig:n dependence}
\end{figure}

In Fig.~\ref{fig:n dependence}, we plot the dependence of the transition curve with respect to the LJ exponent $n$, c.f. \eqref{eq:LJ} changes. We fix $m=6+n$. We observe that as $n$ increases (from $n=1$, in increments of one in the figure), the curve shifts upwards. 

All the results presented in this section correspond to a triangular lattice. The results for square lattices are similar. While the square lattices are equilibrium configurations, for all the curves that generically look like the triangular lattices, the flat state is unstable.

We conclude this section with the following remark on the numerical evaluation of $C$ required to compute the transition curves. 

\begin{remark}
We evaluate the sum over the reciprocal lattice in \eqref{eq:def C} by truncating it over a finite number of lattice points. For all the results in this article, the largest mode number (in both $\mathbf{G}_1$ and $\mathbf{G}_2$ directions) is chosen to be $8$. A convergence study has been summarized in the Supplementary Material.
\end{remark}
\section{Discussion}
\label{sec:discussion}
We discuss the implications of our results  by comparing our analysis to some experimental data. 

\begin{figure}[h!]
    \centering
    \includegraphics[width=\linewidth]{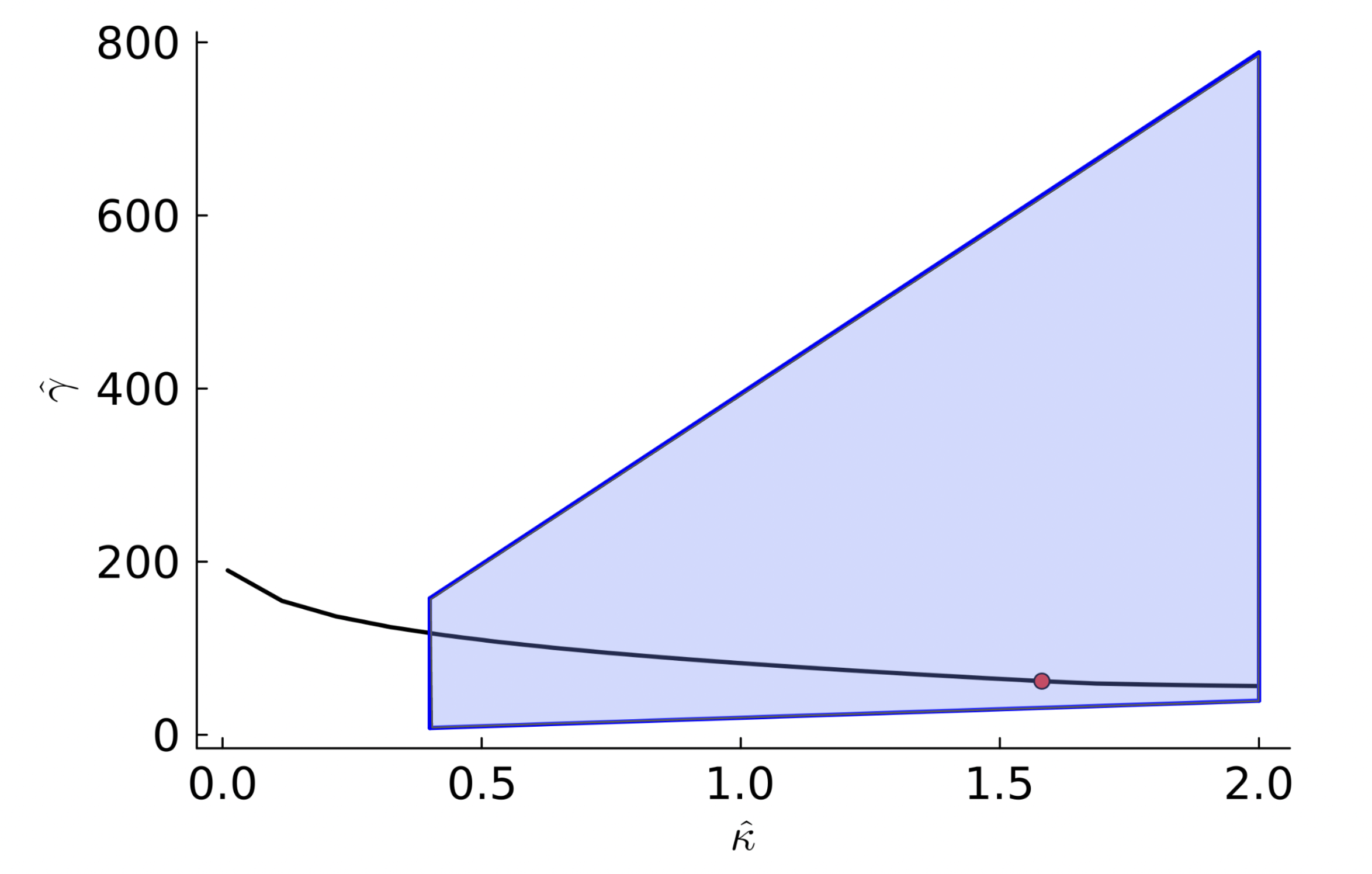}
    \caption{Caption}
    \label{fig:colloid-kappa-gamma}
\end{figure}

For the experiments suggested by Wang et. al.,  \citep{wang2019assembling}, the bending modulus of the DMPC membrane is $\kappa=10k_BT$ (the typical value of $\kappa=20k_BT$ assumes that $E_{bend}=\int\frac{1}{2}\kappa H^2\;da$ in contrast to our expression, c.f., \eqref{eq:energy}). Typical values of surface tension measured in the aforementioned work lies in the range $10^{-3} mN/m$ to $2\times 10^{-2} mN/m$ and the spacing between the colloids (twice the radius) is $r_e=900 nm$. The binding energy ($\epsilon$) is harder to measure. For colloids, this typically falls within the range of $5k_BT$ to $25k_BT$. The macroscopic interaction between the colloids requires a generalized-LJ potential with $n=1$ and $m=7$. These values are chosen by assuming the colloids are spherical and computing the interaction between the two particles, assuming a Derjaguin approximation  \citep{hamaker1937london}. In Fig.~\ref{fig:colloid-kappa-gamma}, we plot the $\hat{\gamma}=r_e^2\gamma/\epsilon$ versus $\hat\kappa = \kappa/\epsilon$ (for $n=1$ and $m=7$). As above, the transition from an $S$-mode to $L$-mode is shown by a red dot. The blue region represents the experimental range of parameters noted above. Recall that for $\hat{\kappa}$ above the transition curve, the flat trivial state is stable, and for configurations below the curve, states on the right of the red dot have the shape of an $S$-mode (with bud-like protrusions). The states left of the red dot have the shape of an $L$-modes (``egg-carton-shaped''). We observe that there is a range of experimental parameter values where a transition from a flat state to either of the mode shapes is possible. 

Note that to generate Fig.~\ref{fig:colloid-kappa-gamma}, we set $\ell=0.62$. We obtain this value by computing the total interaction energy of the colloids (per unit cell) and minimizing it with respect to $\ell$. 

\begin{figure}[h!]
    \centering
    \includegraphics[width=\linewidth]{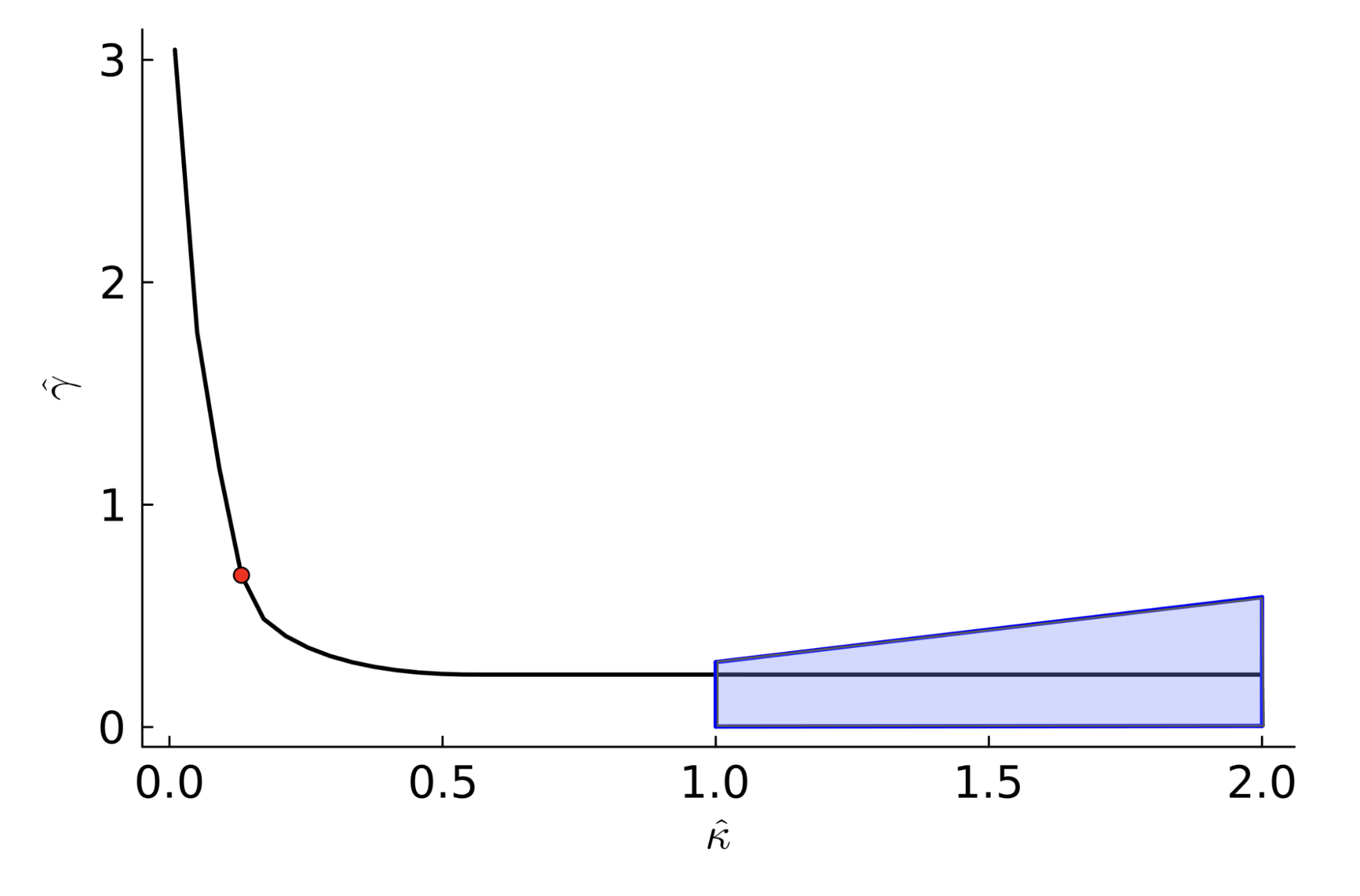}
    \caption{Caption}
    \label{fig:protein-kappa-gamma}
\end{figure}

We explore the implications of our model to the budding of viruses \citep{dharmavaram2019gaussian} and other protein capsids (like clatherine cages  \citep{djakbarova2021dynamic} involved in the endocytosis of molecular cargo). The typical value of surface tension $\gamma$ in the cell is in the range $10^{-3} mN/m$ to $1.2\times 10^{-1} mN/m$  \citep{morris2001cell} and $r_e=10nm$, the typical protein size (Gag protein of HIV, or Clathrin). The binding energy of the proteins must be larger than the thermal energy, and accordingly, we choose $\epsilon$ in the range $5k_BT$ to $10k_BT$. Assuming a standard LJ interaction between proteins, with $n=6$ and $m=12$,  in Fig.~\ref{fig:protein-kappa-gamma}, we plot the transition curve along with the blue region that corresponds to experimental data. We choose $\ell=0.92$, the value obtained by computing the total interaction energy of the colloids (per unit cell) and minimizing it with respect to $\ell$. It is interesting to note that the blue region straddles flat trivial states (above the curve) and budded states of the $S$-mode, below the curve. We predict that for regularly ordered protein assemblies on lipid membranes, it would not be possible to assume the form of an $L$-mode.

During the process of protein-mediated endocytosis, various mechanisms for the generation of curvature in lipid membranes due to proteins have been postulated \citep{djakbarova2021dynamic} e.g., intrinsic curvature of proteins, crowding of proteins, etc.  Our work offers an alternative mechanism for generating curvature. Here, the curvature is generated due to the competition of the bending elasticity of the membrane and inter-protein interaction.

\section{Conclusions}
\label{sec:conclusions}

In this article, we have studied instabilities arising in colloidal crystals on a fluid membrane. 

By applying a Bloch ansatz for the instability modes, we solve the linearized equilibrium equations and determine the conditions under which a flat membrane with particles organized in a regular lattice loses stability. We determine the regions of instability in the $\kappa-\gamma$ plane  (namely, the plane containing values of the bending stiffness and surface tension in the membrane). We find that two modes of out-of-plane instability are possible. One in which the surface assumes a corrugated shape (see Fig.~\ref{fig:L panel}(c)), i.e., the $L$ mode, and the other in which the membrane buckles with bud-like protrusions (see Fig.~\ref{fig:S panel}), i.e., the $S$ mode. We find that the $L$ mode of instability occurs for smaller values of $\hat\kappa$ (and large values of $\hat\gamma$), while $S$ mode occurs for larger values of $\hat\kappa$ (and smaller values of $\hat\gamma$). This is consistent with the expectation that $L$ modes with smaller bending stiffness can support higher curvature, and conversely for $S$ modes. We also compare the predictions of our model with experimental data and provide parameter regions where one would expect the instability to be dominant. 

We also have explored the dependence of the transition curve on other parameters of the model, viz., $\ell$ (related to area per unit cell) and $n$ (exponent of the generalized-LJ potential). We find that the general trend and the presence of two modes of instability are preserved, but the transition curve shits appropriately.

Even though square lattices can be locally stable equilibria for some values of $\ell$ \citep{betermin2015minimization}, we find that any instability from trivial flat state to non-planar modes happens at values of $\ell$ for which the flat state is already unstable. Of course, we have not performed a detailed parameter analysis for the square lattice. We justify this by noting that most experiments only observe a triangular lattice.

It must be noted that in this work, we prove that the flat (trivial) state is stable for parameter values above the transition curve and that as one approaches the transition curve, the flat state loses stability in two possible modes ($S$ and $L$). We do not characterize the stability of these modes. Doing a careful stability analysis of these nontrivial solutions will require a nonlinear analysis which is significantly harder and will be the focus of a future work.

Even though our work focuses on flat membranes, it could be applied to slightly curved membranes, i.e., particle spacing ($\mg{r}$, cf. \eqref {eq:r0})  is small compared to the radius of curvature of the membrane. In this case, the surface tension $\gamma$ is set by the osmotic pressure. When the radius of the curvature is not small compared to the interparticle distance, our results are no longer applicable. The corresponding open problem could be explored in future work.

We conclude by discussing the effects not considered in this work. The adhesion interaction between colloids and the membrane has also been neglected here. Our work has been primarily motivated by the work of Wang et al. \citep{wang2019assembling}, particularly in the high surface tension limit of their experiments. As per their suggestion, the adhesion interaction can be neglected. Incorporating adhesion interactions in our formulation is not as straightforward as the local deformation of the membrane around the particle needs to be accounted for. This means the particles cannot be treated as point particles. Furthermore, the deformation of the particle itself needs to be considered. We have also neglected thermal instabilities and entropic effects, and our analysis assumes quasi-statics. For macroscopic objects like colloids, these effects may be negligible, but applying our results to proteins interactions could be less precise.

Depletion interactions, caused by the solute molecules, are entropy-driven \citep{lekkerkerker2024colloids} and are typically long-range and are always attractive. Since the instabilities considered here are only observed in the repulsive regime (when the particles are under compression), we expect that any depletion effects could delay the onset of the instability. A careful consideration of depletion interaction is challenging because one must consider the forces exerted by two types of solute molecules on the colloidal particle: water molecules (outside the plane of the membrane) and lipid molecules (within the plane of the membrane). While it may be straightforward to consider the effect of water molecules, the effect of lipids, which lie in the plane of the membrane, is still an open problem.

\bibliography{rsc} 
\bibliographystyle{rsc}

\end{document}